%% file: stefan_manuscript.tex
\documentclass[review]{elsarticle}
\makeatletter
\def\ps@pprintTitle{%
 \let\@oddhead\@empty
 \let\@evenhead\@empty
 \def\@oddfoot{\centerline{\thepage}}%
 \let\@evenfoot\@oddfoot}
\makeatother

\pdfoutput=1
\usepackage{graphicx}
\usepackage{amsfonts, amsmath, amssymb}
\usepackage{mathabx}
\usepackage{booktabs,siunitx}
\usepackage[svgnames,table]{xcolor}
\usepackage[tableposition=above]{caption}
\usepackage{pifont}
\usepackage{bm}
\usepackage{cancel}
\usepackage{caption}
\usepackage{color}
\usepackage{enumerate}
\usepackage{float}
\usepackage{hyperref}
\usepackage{soul}
\usepackage[english]{babel}
\usepackage[margin=1in]{geometry}
\usepackage{mathtools}
\usepackage{multirow}
\usepackage{setspace}
\usepackage{subfigure}
\usepackage{stmaryrd}
\usepackage{lineno}
\DeclareUnicodeCharacter{00B3}{\textsuperscript{3}}
\usepackage[ruled,linesnumbered]{algorithm2e}
\RestyleAlgo{ruled}
\SetKwComment{Comment}{/* }{ */}

\SetCommentSty{mycommfont}
\DeclareMathAlphabet{\mathpzc}{OT1}{pzc}{m}{it}

\newcommand \dd[2] {\frac{{\rm d} #1}{{\rm d} #2}}

\renewcommand \d [2]{\frac{{\rm d} #1}{{\rm d} #2}}
\renewcommand \d [1]{{\rm{d}} #1}
\newcommand \D [2]{\frac{\partial #1}{\partial #2}}
\newcommand \DD[2]{\frac{\partial^2 #1}{\partial #2 ^2}}
\newcommand \DDD [2]{\frac{{\rm D} #1}{{\rm D} #2}}
\renewcommand{\vec}[1]{\bm{\mathrm{#1}}}
\newcommand{\V}[1]{\bm{\mathrm{#1}}}

\def \div{\nabla \cdot \mbox{}}
\def \grad{\nabla}

\def \x{\vec{x}}

\def \q{\vec{q}}
\def \n{\vec{n}}

\def \u{\vec{u}}

\def \vD{\vec{D}}

\def \g{\vec{g}}
\def \G{\vec{G}}

\def \Lmu{\vec{L_{\mu}}}

\def \Nx{N_x}
\def \Ny{N_y}

\def \Omegal{\Omega_{\text{l}}}
\def \Omegag{\Omega_{\text{g}}}

\def \f{\vec{f}}

\def \half{\frac{1}{2}}
\def \3half{\frac{3}{2}}
\def \5half{\frac{5}{2}}

\def \mul{\mu^{\text{L}}}
\def \mus{\mu^{\text{S}}}
\def \mug{\mu^{\text{G}}}
\def \n{\vec{n}}

\def \rhol{\rho^{\text{L}}}
\def \rhos{\rho^{\text{S}}}
\def \rhog{\rho^{\text{G}}}
\def \kl{k^{\text{L}}}
\def \ks{k^{\text{S}}}
\def \kg{k^{\text{G}}}

\def \s{\vec{s}}

\def \t{\vec{t}}
\def \u{\vec{u}}

\def \x{\vec{x}}

\def \div{\nabla \cdot \mbox{}}
\def \grad{\nabla}

\def \dx{\Delta x}
\def \dy{\Delta y}

\def \rhos{\rho^{\rm S}}
\def \rhol{\rho^{\rm L}}
\def \rhog{\rho^{\rm G}}
\def \cps{C^{\rm S}}
\def \cpl{C^{\rm L}}
\def \cpg{C^{\rm G}}
\def \Ts{T^{\rm S}}
\def \Tl{T^{\rm L}}

\def \ks{\kappa^{\rm S}}
\def \kl{\kappa^{\rm L}}
\def \kg{\kappa^{\rm G}}

\def \ul{u^{\rm L}}
\def \us{u^{\rm S}}
\def \ug{u^{\rm G}}
\def \el{e^{\rm L}}
\def \es{e^{\rm S}}
\def \ql{\q^{\rm L}}
\def \qs{\q^{\rm S}}
\def \unl{u_{\rm n}^{\rm L}}
\def \uns{u_{\rm n}^{\rm S}}
\def \utl{u_{\rm t}^{\rm L}}
\def \uts{u_{\rm t}^{\rm S}}
\def \pl{p^{\rm L}}
\def \ps{p^{\rm S}}
\def \vul{\vec{u}^{\rm L}}
\def \vus{\vec{u}^{\rm S}}

\def \Omegas{\Omega^{\text{S}}}
\def \Omegal{\Omega^{\text{L}}}
\def \Omegag{\Omega^{\text{G}}}
\def \Omegap{\Omega^{\text{P}}}
\def \alphas{\alpha^{\text{S}}}
\def \alphal{\alpha^{\text{L}}}

\def \mul{\mu^{\text{L}}}
\def \mus{\mu^{\text{S}}}
\def \mug{\mu^{\text{G}}}
\def \pl{p^{\text{L}}}
\def \ps{p^{\text{S}}}
\def \etal{\eta^{\text{L}}}
\def \etas{\eta^{\text{S}}}

\def \Tsol{T^{\rm sol}}
\def \Tliq{T^{\rm liq}}
\def \hsol{h^{\rm sol}}
\def \hliq{h^{\rm liq}}

\newcommand{\upperRomannumeral}[1]{\uppercase\expandafter{\romannumeral#1}}

\newcommand{\REVIEW}[1]{{\color{black}#1}}
\newcommand{\NEW}[1]{{\color{black}#1}}

\begin{document}
\let\today\relax

\begin{frontmatter}
	
\title{A low Mach enthalpy method to model non-isothermal gas-liquid-solid flows with melting and solidification}
\author[SDSU]{Ramakrishnan Thirumalaisamy}
\author[SDSU]{Amneet Pal Singh Bhalla\corref{mycorrespondingauthor}}
\ead{asbhalla@sdsu.edu}

\address[SDSU]{Department of Mechanical Engineering, San Diego State University, San Diego, CA}
\cortext[mycorrespondingauthor]{Corresponding author}

\begin{abstract}
Modeling phase change problems numerically is vital for understanding many natural (e.g., ice formation, steam generation) and engineering processes (e.g., casting, welding, additive manufacturing). Almost all phase change materials (PCMs) exhibit density/volume changes during melting, solidification, boiling, or condensation, causing additional fluid flow during this transition. Most numerical works consider only two phase flows (either solid-liquid or liquid-gas) for modeling phase change phenomena and some also neglect volume/density change of PCMs in the models.  This paper presents a novel low Mach enthalpy method for simulating solidification and melting problems with variable thermophysical properties, including density. Additionally, this formulation allows coupling a solid-liquid PCM with a gas phase in order to simulate the free surface dynamics of PCMs undergoing melting and solidification. We revisit the two-phase Stefan problem involving a density jump between two material phases.  We propose a possible means to include the kinetic energy jump in the Stefan condition while still allowing for an analytical solution. The new low Mach enthalpy method is validated against analytical solutions for a PCM undergoing a large density change during its phase transition. Additionally, a few simple sanity checks are proposed to benchmark computational fluid dynamics (CFD) algorithms that aim to capture the volume change effects of PCMs.
   
\end{abstract}


\end{frontmatter}

\section{Introduction}
The numerical modeling and simulation of phase change materials (PCMs) is a very active area of research because they play a key role in energy systems (e.g., concentrated solar power plants and latent thermal energy storage units~\cite{el2017thermal,allouhi2018optimization,badiei2020performance,hossain2019two,nie2020numerical}), geophysical processes (e.g., sea ice formation and glacier melting~\cite{buffo2021dynamics,buffo2021characterizing}), and manufacturing technologies (e.g., casting, welding, and metal 3D printing~\cite{king2015laser,king2015overview,khairallah2016laser,ly2017metal}). The numerical modeling of PCMs is difficult because the energy equation is nonlinear, and most problems involve liquid flows, and some also involve gas flows and solid motion.   As an example, consider selective laser melting (SLM) of metal powder as PCM for additively printing complex parts. During the SLM process,  a thin powder layer  of thickness 20–100 $\mu$m is deposited with the aid of a roller or a blade, and then fused by a directed laser source that typically scans at a rate of 0.1–1 m/s. After each laser pass over the powder bed, the solid powder particles melt and evaporate, and the molten metal pool solidifies to create the print~\cite{kruth1996basic}. The solidifying metal can entrain gas plumes from the surface, causing porosity and keyhole defects~\cite{karayagiz2019numerical,heeling2017melt,matthews2016denudation,khairallah2016laser} in the finished product. The aforementioned application of metal (powder) as a PCM illustrates the importance of resolving large density gradients (gas-metal density ratio is $\sim10^4$) within multiphysics PCM simulations, which pose stability issues for numerical schemes~\cite{nangia2019robust,pathak20163d,patel2018diffuse}. 

Computational fluid dynamics (CFD) models for simulating the phase change of materials began to be developed in the late 80s and early 90s. According to the way they handle the moving phase boundary, these models can be categorized into two main groups: (\textbf{1})  deforming and (\textbf{2}) fixed grid schemes. Fixed grid schemes offer greater flexibility for incorporating additional physics into heat transfer problems (e.g., fluid flow) and are easier to implement than deforming grid schemes.  They also naturally handle complex topological changes of the interface (merging, pinching, break-up, self-folding), which the deforming grid methods cannot. In the early development of fixed grid numerical methods, the volume change in PCM with melting or solidification was ignored and the variable density in the liquid was described by the Boussinesq approximation. This is still practiced today~\cite{panwisawas2017mesoscale,wu2018numerical,wu2018parametric,gurtler2013simulation,attar2011lattice,panwisawas2017keyhole,aggarwal2018particle}, despite significant improvements in numerical methodologies and the fact that solid–liquid phase changes are always accompanied by a volume change. Since most existing numerical methods can only deal with constant density PCMs, the gas phase is neglected and only melting and solidification are simulated.   In the context of metal AM simulations, a recent review paper by Cook and Murphy~\cite{cook2020simulation} demonstrates that in most of the existing studies, the effect of the surrounding gas flow has been ignored. This is mainly due to the assumption that the powder or substrate cannot move. In reality, surrounding gas flows are prominent near the laser-interaction zone, where extreme thermo-capillary forces can drive gaseous pores out of the melt-pool and/or entrain particles in the melt-pool, resulting in deeper keyhole formation, which has been observed experimentally~\cite{wolff2019situ,ly2017metal}. 

In this paper, a fixed-grid low Mach enthalpy method is developed to capture density change-induced flow during PCM melting and solidification. In this formulation a gas phase is also incorporated and coupled to the solid-liquid PCM region. Our ultimate goal is to develop a simulation method that can handle simultaneous occurrences of evaporation, condensation, melting, and solidification. A method such as this would allow realistic modeling of manufacturing processes such as metal additive manufacturing, in which all four modes of phase change are present at the same time. The original enthalpy method (EM) introduced by Voller and colleagues \cite{voller1987fixed,voller1991eral} for modeling melting and solidification of PCMs assumes the two material phases have the same density. Some recent works~\cite{galione2015fixed,hassab2017effect,dallaire2017numerical,faden2019optimum} have relaxed the constant density requirement of the enthalpy method, but none to our knowledge have explicitly or carefully accounted for the volume change effect in this technique. This becomes even more critical for CFD models \cite{yan2018fully,lin2020conservative,panwisawas2017mesoscale} that consider three phase gas-liquid-solid flows.  Furthermore, existing numerical works have not verified whether using variable thermophysical properties (density, specific heat, thermal conductivity) in the enthalpy method produces an accurate numerical solution when such properties vary widely between phases\footnote{This is primarily due to the fact that in enthalpy methods, any thermophysical property is expressed as a function of a liquid fraction variable that also needs to be solved for. Variable thermophysical properties increase the nonlinearity of enthalpy methods.}. The new low Mach EM is validated using the analytical solution to the two-phase Stefan problem for a PCM that undergoes a substantial volume change during solidification. The two-phase Stefan problem has a non-standard Stefan condition that involves additional jumps in the specific heat and kinetic energy. To derive analytical solutions, the kinetic energy jump term is dropped from the Stefan condition as it is usually small in comparison to latent heat. We discuss a way to retain it in the Stefan condition while still allowing for an analytical solution. 

\section{Jump conditions across the phase-changing interface}\label{sec_jump_conditions}
We follow Myers et al.~\cite{myers2020stefan} to derive the jump conditions across the phase-changing liquid-solid interface. The enthalpy equation and energy equation are distinguished at the end of this section. In addition, we will see how certain jump terms get omitted if the enthalpy equation is used as the starting point for the derivation. To derive the jump conditions, it is convenient to express the governing equations for mass balance, momentum, and energy in conservative form and apply the Rankine-Hugoniot condition across the interface. The conservation laws in differential form\footnote{Alternatively, one can start with the integral form of the conservation principle and derive the jump condition(s); see, for example, Delhaye~\cite{delhaye1974jump}.} read as
\begin{align}
&\D{\rho}{t} + \div (\rho \u ) = 0   \label{eq_mass_cons} \\ 
&\D{\rho \u}{t} + \div (\rho \u \otimes \u + p \mathbb{I}) -  \sigma \mathcal{C} \delta (\x - \s) \n = 0  \label{eq_mom_cons} \\
& \D{}{t}\left(\rho \left[e + \frac{1}{2}| \u|^2 \right] \right) + \div \left( \rho\left[ e + \frac{1}{2} | \u |^2 \right] \u + \q + p \u \right) - \sigma \mathcal{C} \delta (\x - \s) \u \cdot \n = 0.  \label{eq_energy_cons}
\end{align}
In momentum Eq.~\eqref{eq_mom_cons}, $\sigma$ is the surface tension coefficient between the two phases (liquid and solid in this context), $\mathcal{C}$ represents the mean local curvature of the interface, $\s$ represents the position of the interface, $\delta$ is the Dirac delta distribution, and $\n$ is the outward unit normal vector of the interface (pointing outwards from the solid and into the liquid phase in Fig.~\ref{fig_schematic} (A)). In energy  Eq.~\eqref{eq_energy_cons}, $e$ denotes the internal energy and $\q = -\kappa \grad T$ is the conductive heat flux. Integrating Eqs.~(\ref{eq_mass_cons})-(\ref{eq_energy_cons}) across a finite region around the liquid-solid interface and then letting the volume of the region tend to zero, gives  

\begin{enumerate}
\item  Mass jump across the interface:    
\begin{align} 
(\rhol - \rhos)u^*  = (\rhol \vul - \rhos \vus) \cdot \n = (\rhol \unl - \rhos \uns).  \label{eq_mass_jump}
\end{align}
Here, $u^* = \u \cdot \n$ represent the normal velocity of the interface, and $\unl = \vul \cdot \n$ and $\uns = \vus \cdot \n$ represent the near-interface normal component of velocity of the liquid and solid phase, respectively. The following relation can be obtained by taking the velocity in the solid to be zero $\vus = 0$ 
\begin{equation}
\unl = \left( 1 - \frac{\rhos}{\rhol} \right) u^*. \label{eq_ul_u*} 
\end{equation}

\item  Momentum jump across the interface:   
\begin{equation}
(\rhol \vul - \rhos \vus)u^* = \left(\rhol \vul \otimes \vul + \pl \mathbb{I} \right) \cdot \n - \left(\rhos \vus \otimes \vus + \ps \mathbb{I} \right) \cdot \n + \sigma  \mathcal{C} \n.  \label{eq_mom_jump}
\end{equation}
The above vector equation can be expressed in terms of normal and tangential jumps. Taking an inner product  of Eq.~\eqref{eq_mom_jump} with the unit normal vector $\n$  yields 
\begin{align} 
(\rhol \unl - \rhos \uns) u^* &= \rhol (\unl)^2  - \rhos (\uns)^2  + \pl - \ps  + \sigma  \mathcal{C} \nonumber \\
\hookrightarrow \;  \ps - \pl &= \rhol \unl (\unl - u^*) + \sigma \mathcal{C} = -\rhos \unl u^* + \sigma \mathcal{C}, \label{eq_normal_mom_jump}
\end{align}
in which we used Eq.~\eqref{eq_mass_jump} to express pressure jump in terms of $\rhos$.  Likewise, jump in the tangential momentum across the interface is obtained by taking the inner product of Eq.~\eqref{eq_mom_jump} with the unit tangent vector $\t$
\begin{align} 
(\rhol \utl   - \rhos \uts) u^* &= \rhol \unl \utl    - \rhos \uns \uts  \nonumber \\
\hookrightarrow \; (\rhol - \rhos) u^* &= \rhol \unl - \rhos \uns. \label{eq_tangential_mom_jump}
\end{align}
Here, $\utl = \vul \cdot \t$ and $\uts = \vus \cdot \t$ denote the tangential velocity component of the liquid and solid phase, respectively. In addition, we invoked the no-slip condition at the interface to equate the tangential velocities of the two phases, $\utl = \uts$. The no-slip assumption makes the mass and tangential momentum jumps equivalent.  

\item  Energy jump across the interface:   
\begin{align} 
\rhol \left( \el + \frac{ |\vul|^2 }{2}\right) (u^* - \unl) - \rhos \left( \es + \frac{ |\vus|^2 }{2}\right) (u^* - \uns) &= [(\ql - \qs) + (\pl \vul - \ps \vus)] \cdot \n   + \sigma \mathcal{C} u^* \nonumber  \\
\hookrightarrow \;  \left[ \rhos \left( \el  - \es + \frac{ |\vul|^2 }{2}\right) - \sigma \mathcal{C} - \left( 1 - \frac{\rhos}{\rhol}\right) \pl \right] u^* &= [(\ql - \qs)] \cdot \n, \label{eq_energy_jump}    
\end{align}
in which Eqs.~(\ref{eq_mass_jump}) and (\ref{eq_ul_u*}) have been used to simplify some terms.   
\end{enumerate}  

The Stefan condition is obtained from the energy jump Eq.~\eqref{eq_energy_jump} by expressing internal energy in terms of temperature $T$ and latent heat $L$
\begin{align}
\es &= \cps(\Ts - T_r) - \frac{\ps}{\rhos},  \label{eq_es} \\
\el &= \cpl(\Tl - T_r) + L - \frac{\pl}{\rhol}. \label{eq_el}
\end{align}
Substituting Eqs.~\eqref{eq_es} and \eqref{eq_el} into Eq.~\eqref{eq_energy_jump} and using Eqs.~\eqref{eq_normal_mom_jump} and  \eqref{eq_mass_jump} for further simplifications, yields the Stefan condition
\begin{equation}
\rhos \left[(\cpl - \cps)(T_m - T_r) + L-\frac{1}{2}\left(1- \left( \frac{\rhos}{\rhol} \right)^2\right) (u^*)^2\right]u^* =   [\ks \grad \Ts   -\kl \grad \Tl]\cdot \n. \label{eq_app_stefan}
 \end{equation}
 Here, $T_m$ denotes the melting/solidification temperature attained by the two phases at the interface and $T_r$ is the phase change temperature in the bulk measured at a reference pressure. 

Note that the Stefan condition (Eq.~\eqref{eq_app_stefan}) is derived from the energy jump Eq.~\eqref{eq_energy_jump}, which considers the change in kinetic energy between liquid and solid across the interface. Several textbooks derive the Stefan condition from the enthalpy equation and do not include jumps related to kinetic energy. The reason for this can be traced back to the derivation of the enthalpy equation, which is determined by subtracting the kinetic energy equation from the energy equation. The subtraction eliminates terms like $\frac{1}{2}|\u|^2$. The kinetic energy equation is obtained by taking an inner product of velocity with the momentum equation and manipulating derivatives of velocity and pressure to obtain terms like $\div(\frac{1}{2}|\u|^2)$ and $\div(p\u)$. It is not possible to manipulate derivatives for a phase change problem since both velocity and pressure are discontinuous across the interface. 

\section{Analytical solution to the two phase Stefan problem with density change} \label{sec_analytical_stefan}

 \begin{figure*}
\centering
\includegraphics[width=0.7\linewidth]{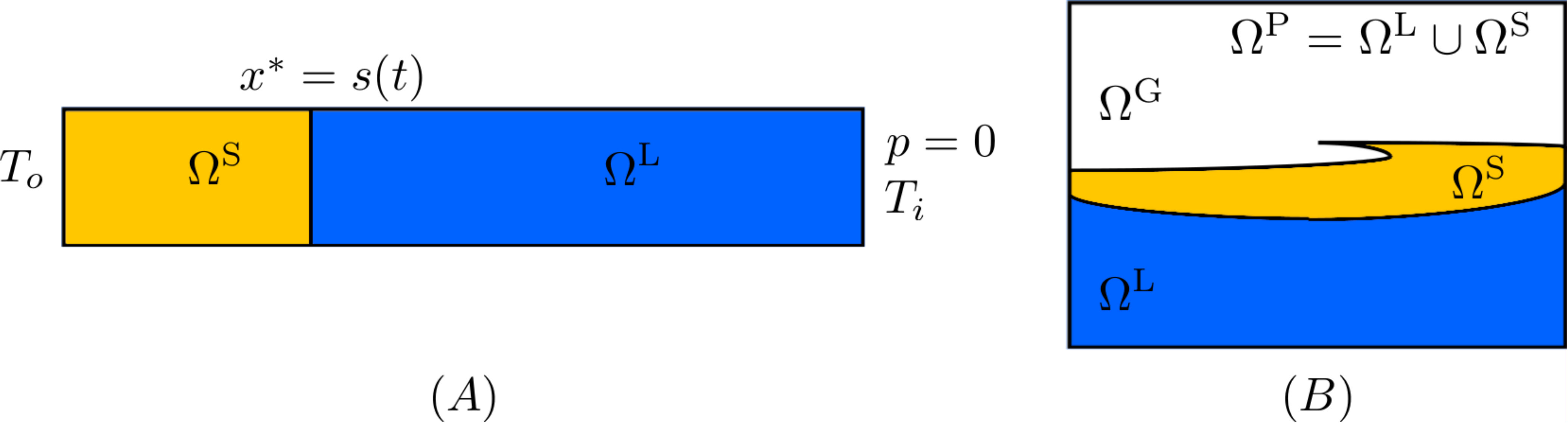}
\caption{Schematics of the (A) two-phase and (B) three-phase problems examined in this study. A liquid phase is represented by blue, a solid phase by yellow, and a gas phase by white. Gas and PCM regions are tracked using the Heaviside function $H$, which is defined to be 1 in the PCM domain and 0 in the gas domain. Liquid and solid phases are tracked by the liquid fraction variable $\varphi$, which is equal to 1 in the liquid phase and 0 in the solid phase. A mushy region is defined by $0 < \varphi < 1$.}
\label{fig_schematic}
\end{figure*}

In this section, we revisit the two-phase Stefan problem involving a density jump from Alexiades and Solomon's textbook~\cite{alexiades2018mathematical}. Unfortunately, despite its simplicity, this problem has not received much attention in the CFD literature. Meanwhile, the single phase Stefan problem, where only the heat equation is used and no density change-induced fluid flow is involved, remains the gold standard for validating advanced CFD algorithms for modeling melting and solidification \cite{huang2022consistent,yan2018fully,javierre2006comparison} and boiling and condensation \cite{gibou2007level,khalloufi2020adaptive} phenomena. In their textbook~\cite{alexiades2018mathematical}, the authors drop the kinetic energy jump term from the Stefan condition because it ``destroys" the similarity solution. As we show  in this section, this is not the case. A method for including it in the analytical solution is discussed. In the early stages of solidification, the interface velocity is infinite and the kinetic energy term dominates. Likewise, when a cylindrical or spherical material melts, its melting rate tends to infinity towards the end of the process. There are situations (certain time periods and length scales) where the kinetic energy term becomes important, and our analytical solution may prove useful in those instances. 

Consider solidification of a liquid in a large static domain $\Omega := 0 \le x \le l$ that is closed on the left and open on the right. The same setup and analytical methodology can also be applied to the evaporation/condensation problem. Liquid occupies the whole domain at $t = 0$ and has a uniform temperature $T_i$ greater than that of solidification temperature $T_m$. The temperature at the left boundary ($x=0$) is suddenly lowered to $T_o < T_m$ at $t = 0^+$, which remains constant thereafter. The solidification front having position $x^* = s(t)$ moves in the positive $x$-direction as shown in Fig.~\ref{fig_schematic}(A).  The thermophysical properties, i.e., density, specific heat, and thermal conductivity of the liquid phase are denoted $\rhol$, $\cpl$, and $\kl$, respectively. The respective quantities for the solid phase are denoted $\rhos$, $\cps$, and $\ks$. Solid-to-liquid density ratio is denoted $R_\rho = \rhos/\rhol$. In the case where $R_\rho < 1$, the liquid expands as it solidifies, causing an additional flow in the direction of solidification. Alternatively, if $R_\rho > 1$, the fluid shrinks as it solidifies, causing a flow in the opposite direction. We seek analytical solutions for temperature, velocity, and pressure in the liquid and solid domains. Starting with the Rankine-Hugoniot relation for the mass balance equation (see Eq.~\eqref{eq_mass_jump}) the jump in velocity across the interface is obtained    
\begin{equation}
\llbracket u \rrbracket =  \ul (s^{+},t )- \us(s^{-}, t) =  \ul (s^{+},t ) = (1-R_\rho) \dd{s}{t}. \label{eqn_vel_jump}
\end{equation}
Here, $s^{+}$ and $s^{-}$ represent spatial locations just ahead and behind the interface, and $u^* = \d s / \d t$ represents interface speed.  The velocity in the solid domain $\Omegas(t) := 0 \le x < s(t)$ is taken to be zero\footnote{In the case of evaporation and condensation, the solid phase will be replaced by gas/vapor in the analytical model. In the vapor phase, the velocity is also zero everywhere $\ug \equiv 0$. It follows from the continuity equation $\partial \ug/ \partial x=0$ and the zero-velocity boundary condition imposed at the left end ($x = 0)$.}, i.e., $\us(\Omegas,t) \equiv 0$, while in the liquid domain $\Omegal(t) := s(t) < x \le l$, it is uniform as per the continuity equation $\partial \ul/ \partial x= 0$. Therefore, liquid velocity can be obtained directly from interface speed as $\ul(\Omegal,t) \equiv  \ul (s^{+},t )$. Once the solid and fluid velocities have been determined, it is possible to solve the phase change problem with volume change effects by solving the energy equation for both solid and liquid phases

\begin{align}
\rhos \cps \left( \frac{\partial \Ts}{\partial t} + \us \frac{\partial \Ts}{\partial x} \right) &=\ks\frac{\partial^2 \Ts}{\partial x^2} \, \in  \Omegas(t),  \label{eq_temp_solid}  \\
\rhol \cpl \left( \frac{\partial \Tl}{\partial t} + \ul \frac{\partial \Tl}{\partial x} \right)&=\kl\frac{\partial^2 \Tl}{\partial x^2} \,  \in \Omegal (t), \label{eq_temp_liquid}   
\end{align}
by taking $\us(x,t) \equiv 0$ and $\ul (x,t) \equiv \ul (s^{+},t )$. Five boundary conditions are required to determine $\Ts(x,t)$, $\Tl(x,t)$ and $x^*= s(t)$ completely. These include two boundary conditions  $\Ts(0,t) = T_o$ and $\Tl(l, t) = T_i$ and three interfacial conditions:
\begin{align}
&\Ts(x^*, t) = \Tl(x^*, t) = T_m, \label{eq_interface_temp}\\
&\rhos\left[(\cpl - \cps)(T_m - T_r) + L-\frac{1}{2}(1-R_\rho^2) \left( \dd{s}{t} \right)^2\right] \dd{s}{t}    \nonumber \\ 
& = \left( \ks\frac{\partial \Ts}{\partial x}  -\kl \frac{\partial \Tl}{\partial x}  \right)_{x^*}.  \label{eq_stefan_condition}
\end{align}
Here, $T_r$ is the bulk phase change temperature measured at a specified reference pressure $p_r$, and $L$ is the latent heat of melting/solidification. Eq.~\eqref{eq_stefan_condition}, which is commonly referred to as the Stefan condition, contains terms related to latent heat and jumps in the specific heat, density, and kinetic energy of the two material phases; see Sec.~\ref{sec_jump_conditions}.

As shown in the Appendix Sec.~\ref{sec_similarity_soln}, Eqs.~\eqref{eq_temp_solid} and \eqref{eq_temp_liquid} admit similarity solution of the form   
\begin{align}
\Ts(x,t) &= T_o + A(\lambda(t))\,\text{erf}\left(\frac{x}{2\sqrt{\alphas t}}\right), \label{eq_Tprofile_solid} \\ 
\Tl(x,t) &=  T_i + B(\lambda(t))\,\text{erfc}\left(\frac{x}{2\sqrt{\alphal t}}- \frac{s(t)}{2\sqrt{\alphal t}}(1-R_\rho)\right), \label{eq_Tprofile_liquid}
\end{align} 
 in which  $\alphas = \ks/(\rhos \cps)$ and   $\alphal = \kl/(\rhol \cpl)$ are the solid and liquid thermal diffusivities, respectively, and $\lambda(t) = \frac{s(t)}{2\sqrt{\alphal t}}$ is a yet to be determined function of time. These temperature profiles satisfy the boundary and initial conditions of their respective equations.  {The unknown functions $A$ and $B$ appearing in the temperature profiles $\Ts(x,t)$ and $\Tl(x,t)$ implicitly depend on time through $\lambda(t)$. If $\lambda(t)$ is time-(in)dependent, so are $A$ and $B$. To our knowledge, all textbooks and papers first assume that $A$ and $B$ are constants, and then deduce that $\lambda$ has to be a time-independent constant. In contrast, we posit that $A$ and $B$ should actually be determined the other way around.}  To wit, the interface temperature condition written in Eq.~\eqref{eq_interface_temp} 
\begin{equation*}
T_o + A(\lambda(t))\,\text{erf}\left(\frac{s(t)}{2\sqrt{\alphas t}}\right) = T_i + B(\lambda(t))\,\text{erfc}\left(\frac{s(t)}{2\sqrt{\alphal t}} R_\rho \right) = T_m,
\end{equation*}
allows us to express $A$ and $B$ as a function of $\lambda(t)$:
\begin{equation}
A(\lambda(t)) = \frac{T_m-T_0}{\text{erf}\left( \lambda(t) \sqrt{\frac{\alphal}{\alphas}} \right) } \qquad \text{and}   \qquad B(\lambda(t)) =  \frac{T_m-T_i}{\text{erfc}\left(\lambda(t) R_\rho\right)}.
\end{equation}  
 The temperature distribution in the solid phase is therefore
\begin{equation}
\Ts = T_o + (T_m-T_o)  \text{erf}\left(\frac{x}{2\sqrt{\alphas t}}\right) \bigg/ \text{erf} \left(\lambda(t) \sqrt{ \frac{\alphal}{\alphas}}\right)
\label{eq_temp_solid2}
\end{equation}
and in the liquid phase is
\begin{equation}
\Tl = T_i + (T_m-T_i) \frac{\text{erfc}\left(\frac{x}{2\sqrt{\alphal t}}-\lambda(t)\left(1-R_\rho\right)\right)}{\text{erfc}\left(\lambda(t) R_\rho\right)}.
\label{eq_temp_liquid2}
\end{equation}
By substituting Eqs.~\eqref{eq_temp_solid2}, \eqref{eq_temp_liquid2}, and $s(t) = 2\lambda(t)\sqrt{\alphal t}$
into the Stefan condition (Eq.~\eqref{eq_stefan_condition}), we obtain a governing equation for $\lambda(t)$ \footnote{Due to the absence of the solid phase at the beginning, $\Ts(x,t = 0)$ is not defined. Furthermore, since $\lambda$ is derived using the solutions to $\Ts(x,t)$ and $\Tl(x,t)$, it is defined for $t > 0$. This can also be argued from the definition of interface position $s(t) = 2\lambda(t)\sqrt{\alphal t}$, which does not exist at $t = 0$. On the other hand $\Tl(x,t = 0)$ is defined and is independent of $\lambda$ thanks to the erfc(.) function in the numerator of Eq.~\eqref{eq_temp_liquid2}, which approaches zero as $t \rightarrow 0$.} 
\begin{align}
& \rhos \left[ L^{\rm eff}-\frac{(1-R_\rho^2)}{2}\left(\frac{\lambda^2 \alphal}{t}\right)\right]\lambda \sqrt{\alphal} =  \nonumber \\ 
& \ks \frac{T_m-T_o} {\text{erf}\left(\lambda  \sqrt{\frac{\alphal}{\alphas}} \right)} \frac{e^{-\lambda^2 \alphal / \alphas }}{\sqrt{\pi \alphas}}+ \kl\frac{T_m-T_i}{\text{erfc}\left(\lambda R_\rho\right)} \frac{e^{-\lambda^2 R_\rho^2}}{\sqrt{\pi \alphal}}  + \mathcal{O}\left( \left(\frac{\d \lambda}{\d t}\right)^3,.. \right),
\label{eq_lambda_differential}
\end{align} 
in which $L^{\rm eff} = L + (\cpl - \cps)(T_m - T_r)$ is the effective latent heat. In the equation above $\mathcal{O}(...)$ contains terms related to the time derivatives of $\lambda(t)$, which arise from differentiating $s(t)$ 
\begin{equation}
\dd{s}{t} = \lambda \sqrt{\frac{\alphal}{t}} +  2\sqrt{\alphal t} \; \dd{\lambda}{t}. \label{eq_dsdt}
\end{equation}
At the beginning of the solidification process, the interface speed $\d s/\d t \rightarrow \infty$ due to the leading-order $1/\sqrt{t}$ term. By retaining only this term for $\dd{s}{t}$ in the Stefan condition, we obtain a time-dependent transcendental equation for $\lambda(t)$ instead of a differential one
\begin{align}
& \rhos \left[ L^{\rm eff}-\frac{(1-R_\rho^2)}{2}\left(\frac{\lambda^2 \alphal}{t}\right)\right]\lambda \sqrt{\alphal} =  \nonumber \\ 
& \ks \frac{T_m-T_o} {\text{erf}\left(\lambda  \sqrt{\frac{\alphal}{\alphas}} \right)} \frac{e^{-\lambda^2 \alphal / \alphas }}{\sqrt{\pi \alphas}}+ \kl\frac{T_m-T_i}{\text{erfc}\left(\lambda R_\rho\right)} \frac{e^{-\lambda^2 R_\rho^2}}{\sqrt{\pi \alphal}}.
\label{eq_transcendental}
\end{align} 
\REVIEW{Furthermore, we demonstrate that the $\d \lambda/ \d t$ terms can be omitted from Eq.~\eqref{eq_transcendental} through numerical experiments in Sec.~\ref{sec_sol_trans_eqn}.}

{From Eq.~\eqref{eq_transcendental} (and also Eq.~\eqref{eq_lambda_differential}), it is clear that if we keep the density/kinetic energy jump term in the Stefan condition, $\lambda$ is an explicit function of time, so $A(\lambda)$ and $B(\lambda)$ are implicit functions of time. By dropping the density/kinetic energy jump term from the Stefan condition, which is done in the literature, $\lambda$ becomes independent of time. In this situation, $A(\lambda)$ and $B(\lambda)$ are constants. }   


The closed form solution for $\Ts(x,t)$, $\Tl(x,t)$, and $\ul(x,t)$ can be written once $\lambda(t)$ has been found from Eq.~\eqref{eq_transcendental} (without considering $\d \lambda / \d t$ terms). When $R_\rho=1$ and $L^{\rm eff} = L$, the present analytical solution reduces to the standard Stefan problem solution given in the Hahn and \"{O}zi\c{s}ik textbook~\cite{hahn2012heat}.


Next we find the variation of pressure inside the fluid and solid phases. First consider the fluid momentum equation 
\begin{align}
\rhol  \left( \frac{\partial \ul}{\partial t} + \ul \frac{\partial \ul}{\partial x}  \right) &= - \frac{\partial \pl} {\partial x} + \mul \frac{\partial^2 \ul}{\partial x^2},  \label{eq_liquid_mom}
\end{align}
which simplifies to  
\begin{equation}
\rhol  (1 - R_\rho) \frac{\d^2 s}{\d t^2} = -\frac{\partial \pl} {\partial x}
\end{equation} 
when $\ul \equiv (1-R_\rho) \frac{\d s}{\d t}$ is substituted in Eq.~\eqref{eq_liquid_mom}.  Integrating the above equation yields a linear variation of pressure within $\Omegal(t)$: $s(t) < x \le l$ as
\begin{align}
\pl (x,t) &= p_i - \frac{\lambda(t)}{2\sqrt{t^3}}  (l-x) \left(\rhol - \rhos\right) \sqrt{\alphal}. \label{eq_pl}
\end{align}   
Here, $p_i$ is the pressure of the liquid phase at the right end ($x = l$). Due to the assumption that the liquid phase is incompressible, the domain length $l$ needs to be finite. $l$, however, is large enough to accommodate the plateau region of the error function used in Eq.~\eqref{eq_Tprofile_liquid}.  In the solid phase, we obtain a uniform pressure by considering the momentum equation with $\us \equiv 0$
\begin{align}
\ps (x,t) &= \pl (s^+,t) -  \llbracket p \rrbracket    \nonumber \\ 
& =  p_i - \frac{\lambda(t)}{2\sqrt{t^3}} (l-s(t)) \left(\rhol - \rhos\right) \sqrt{\alphal} -  \llbracket p \rrbracket,  \label{eq_ps}
\end{align}  
in which  $\llbracket p \rrbracket =   \pl (s^{+},t )- \ps(s^{-}, t) = \rhos (1-R_\rho) \left(\frac{\d s}{\d t}\right)^2 $ is the jump in pressure across the interface; see Eq.~\eqref{eq_normal_mom_jump}.  


This completes the analytical derivation of the Stefan problem involving jumps in liquid and solid thermophysical properties and we now proceed to the new low Mach enthalpy method formulation.   

\section{A low Mach enthalpy method}    \label{sec_lm_em}

The fixed-grid CFD techniques for modeling phase change phenomena can be divided into two main categories: (\textbf{1}) sharp and (\textbf{2}) diffuse interface methods. A sharp technique treats the phase boundary as an infinitesimally thin surface, whereas a diffuse technique smears it over a few grid cells. Therefore, the former class of methods can explicitly impose the jump conditions of the governing equations within the solution methodology. Using a diffuse interface formulation, the phase transition occurs across a finite ``mushy" region, and thus there is no jump in the governing equations as all quantities vary continuously. In spite of this, diffuse interface methods are quite popular in the literature due to their simplicity of implementation, robustness, and ability to handle more than two phases simultaneously. We consider the diffuse interface method for simulating solidification/melting of a PCM in this study. In addition, the proposed method allows us to couple a solid-liquid PCM with a passive gas phase without posing major problems. 

Specifically, we consider the enthalpy method (EM) pioneered by Voller and colleagues \cite{voller1987fixed,voller1991eral} for modeling melting and solidification of PCMs. Phase field methods (PFM) are another popular diffuse domain approach for modeling phase change phenomena \cite{boettinger2002phase,huang2022consistent}. A major advantage of EM over PFM is that, unlike PFM, the EM does not require additional material parameters (such as mobility, Gibbs-Thompson, linear kinetic coefficients, mixing energy density, double-well potential function, etc.) that are usually empirically selected during numerical simulation. The solid-liquid interface is implicitly tracked in the EM using the liquid fraction variable $\varphi (\x, t)$ that is defined over the entire domain ($\x$ denotes a spatial location in $\Omega$). $\varphi$ is defined as 1 in the liquid phase, 0 in the solid phase, and between 0 and 1 in the transition/mushy zone. A temperature interval of $\Delta T=\Tliq- \Tsol$ (where $\Tliq$ represents the liquidus temperature at which solidification begins, and $\Tsol$ represents the solidus temperature at which full solidification occurs) is chosen to represent the range over which the phase change occurs. This assumption is based on the fact that for metal alloys and glassy substances there is no single melting temperature $T_m$ because the phase change occurs over an extended range of temperatures from $\Tsol$ to $\Tliq$, and there is a mushy zone between the all solid and all liquid regions \cite{hahn2012heat}. The energy/enthalpy equation implicitly models phase change---upon changing phase, the grid cell's enthalpy adjusts to account for latent heat release or absorption, which in turn changes the $\varphi$ value. EMs are typically implemented as source-based methods \cite{voller1991eral}, where the material enthalpy is divided into sensible and latent components. Newton's method is used to solve the nonlinear energy equation containing latent heat as a source term.  Many CFD softwares, including ANSYS Fluent and OpenFOAM, support source-based EM (SB-EM). Studies using SB-EM have often ignored the phase change induced fluid motion resulting from a density jump between liquid and solid phases. Where applicable, the Boussinesq approximation is used in the momentum equation to account for density variations within the liquid phase (e.g., to model natural convection in a melting PCM~\cite{hu1996mathematical}). 

Some authors have only recently begun considering solid and liquid densities differently when utilizing the EM method~\cite{galione2015fixed,hassab2017effect,dallaire2017numerical,faden2019optimum}. To solve variable-density mass, momentum, and energy equations, these works have employed classical finite volume algorithms such as SIMPLE and PISO \cite{moukalled2016finite,patankar2018numerical}. These works suffer from the following shortcomings:
\begin{enumerate}

\REVIEW{ \item While the density field is defined by the liquid-fraction variable $\varphi$, i.e., an equation of state (EOS) is defined for the system, it is not explicitly used to constrain the velocity field. In other words, the continuous formulation does not explicitly distinguish between the bulk of phases with no change in material volume (where velocity is divergence-free) and the narrow mushy region that allows changes in material volume (where velocity is not divergence-free).  The previous formulations also do not guarantee that when all of the liquid has solidified, or when all of the solid has melted, there are no further volume changes in the system.  } 

\REVIEW{ \item A temperature equation is derived from the enthalpy equation. It is done by expressing enthalpy as a function of temperature, for example, by using $h = C (T - T_r)$ type of relations. Here, $h = e + p/\rho$ denotes the specific enthalpy. However, this conversion has a disadvantage in that the specification of specific heat $C$, that depends on $\varphi$, becomes ambiguous. This is because $\varphi$  evolves with $T$ and it must also be solved for along with temperature. When solving for $T$ and $\varphi$, $C$ is usually held constant in the numerical implementation. As a result, the $h = C (T - T_r)$ relation can be satisfied only weakly. }

{  \item The advection of temperature or enthalpy involves a mass flux term $\V{m_\rho} = \rho \u$. The discrete versions of mass and energy fluxes must be strongly coupled to each other to ensure the numerical stability of high density ratio flows. Prior works had not ensured this coupling.    }
    
{ \item Lastly, the prior numerical algorithms are only qualitatively validated against complex experiments \cite{Tan09,Beckermann88} (in complicated geometries and configurations). This includes qualitatively comparing the interface evolution between simulations and experiments \cite{Tan09,galione2015fixed}.  As a consequence, it is unclear how well the prior continuous and discrete formulations capture density-induced flows or volume changes in PCMs.} 

\end{enumerate}


\REVIEW{To overcome the aforementioned shortcomings in the prior works, we re-formulate the original enthalpy method as a low Mach technique. A low Mach approach has been traditionally applied to gas dynamics applications, like combustion \cite{pember1998adaptive,hosseini2022low} and astrophysical flows \cite{bell2004adaptive,gilet2013low}, however it can also be applied to fluid flow problems. Bell, Donev, and colleagues used a low Mach formulation to simulate multispecies liquid flows at mesoscales \cite{donev2014low, donev2015low, nonaka2015low}. Our new low Mach EM formulation keeps the flow velocity divergence-free (div-free for short) in the bulk of solid and liquid phases. In the narrow mushy region between solid and liquid, where the material volume changes, the velocity is not divergence-free (non-div-free for short). The conditions on the velocity field are made explicit in both the continuous and discrete versions of our low Mach EM. A minor change in the EOS allows us to include a (passive) gas phase in the original solid-liquid PCM system. We assume that gas is incompressible and the formulation ensures that the gas domain's velocity is div-free. We solve the nonlinear enthalpy equation as it is without dividing it into sensible and latent components, i.e., we do not follow the source-based approach. This makes the method more general and allows us to incorporate the gas phase more easily. Furthermore, this avoids the ambiguity of defining specific heat in the domain when solving for enthalpy and liquid fraction. We also solve an additional mass balance equation to strongly couple mass advection with energy and momentum advection. For high density ratio flows, this step ensures that the numerical scheme remains stable. }The new low Mach EM is validated using the analytical solution to the two-phase Stefan problem for a PCM that undergoes a substantial volume change during solidification. To illustrate the practical utility of our formulation, we simulate a metal casting problem showing a pipe defect \cite{stefanescu2015science} caused by the volume shrinkage of solidifying metal. Pipe defects are captured only when the velocity field is non-div-free, i.e., they cannot be captured by relying solely on a variable density field in the momentum and energy equations.    

\subsection{Mathematical formulation}   \label{sec_math_formulation}
 
In our technique the gas-PCM interface $\Gamma(t)$ is tracked using the zero-contour of the signed distance function $\phi (\x,t)$. $\phi$ is defined to be positive in the PCM region $\Omegap(t) = \Omegas(t) \cup \Omegal(t)$ and negative in the gas region $\Omegag(t)$. It is advected with the non-div-free velocity  $\u(\x,t)$ 
\begin{equation}
\DDD{\phi}{t} = \D{\phi}{t}+\u\cdot\grad{\phi}=0. 
\label{eq_level_set}
\end{equation}
A smoothed Heaviside function $H(\x,t)$ is used in conjunction with the SDF $\phi$ to distinguish the gas and PCM regions: $H$ takes a value of 0 in the gas region, 1 in the solid-liquid PCM region, and smoothly transitions from 0 to 1 around $\Gamma = \Omegag \cap \Omegap$ with a prescribed width of 2 grid cells on either side of the gas-PCM interface; see Fig.~\ref{fig_schematic}(B).  
Using the Heaviside function $H$ and the liquid fraction variable $\varphi$ any thermophysical property $\beta$ (e.g., $\rho$, $C$, $\kappa$) can be uniquely defined throughout the domain   
\REVIEW{
\begin{subequations} 
\begin{alignat}{2}
& \beta = \beta^{\rm{G}}+( \beta^{\rm{S}}- \beta^{\rm{G}})H+( \beta^{\rm{L}}- \beta^{\rm{S}})H\varphi.  \label{eq_material_properties}  \\
& \rho= \rhog+( \rhos- \rhog)H+( \rhol- \rhos)H\varphi.  \label{eq_EOS}
\end{alignat}
\end{subequations} 
When $\beta = \rho$, we get the EOS written in Eq.~\eqref{eq_EOS}. 

 The EOS and the mass balance equation  provide a kinematic constraint on the velocity field
\begin{align}
& \D{\rho}{t}+\div{\left(\rho \u\right)}=0, \nonumber \\
 \hookrightarrow & \D{\rho}{t}+\rho \div{\u} + \u \cdot \grad{\rho}=0, \nonumber \\
\hookrightarrow & \div{\u}=-\frac{1}{\rho}\left(\D{\rho}{t}+\u \cdot \grad{\rho}\right)=-\frac{1}{\rho}\DDD{\rho}{t}.
\label{eq_mass_blance}
\end{align}
The velocity divergence constraint can be expressed in terms of liquid fraction $\varphi$ and Heaviside function $H$ using the EOS (Eq.~\eqref{eq_EOS}) as 
}
\begin{align}
\div \u & = -\frac{1}{\rho}\DDD{\rho}{t} \nonumber \\
	&  =  -\frac{1}{\rho} \left( (\rhos - \rhog) \DDD{H}{t} + (\rhol-\rhos)\DDD{\left(H \varphi\right)}{t} \right)  \nonumber \\
           & =  -\frac{1}{\rho} \left( (\rhos - \rhog) \DDD{H}{t} + (\rhol-\rhos)\left(H\DDD{\varphi}{t} + \varphi \DDD{H}{t}\right) \right)  \nonumber \\
           & = \frac{(\rhos-\rhol)}{\rho} H \DDD{\varphi}{t}. 
\label{eq_divu}
\end{align}
In~Eq.~\eqref{eq_divu} we have used $\DDD{H}{t} = 0$ as $H$ follows the same linear advection equation (Eq.~\eqref{eq_level_set}) as $\phi$.   \REVIEW{Having derived the low Mach Eq.~\eqref{eq_divu} above, we now provide a physical rationale for the low Mach formulation of the enthalpy method. 

Since solid and liquid phases are assumed to be incompressible, the characteristic sound speed is infinite in both solid and liquid regions. This means that in the bulk of both phases, the Mach number of the flow is zero. The mushy region between all solid and liquid phases is a very narrow area that is of the order of a few atomic/molecular diameters. Consequently, the characteristic sound speed in the mushy region is expected not to deviate significantly from the bulk solid and liquid phases, and it remains close to infinity. This \textit{ansatz} allows us to employ a low Mach model to express density as a function of liquid fraction, which in turn is a function of enthalpy. A derivation of $\varphi$-$h$ relation will be provided in this section. Low Mach models also imply that variations in density don't affect thermodynamic pressure $\tilde{p}$. Additionally, the pressure variable $p$ which appears in the momentum equation is mechanical in origin. It serves as a Lagrange multiplier that enforces the kinematic constraint on the velocity field as written in Eq.~\eqref{eq_divu}. We remark that although we call the new EM  a ``low Mach" method, it is actually a zero Mach method. This is the common name for the class of models described by equations such as~\eqref{eq_material_properties} and~\eqref{eq_divu}. It is similar to how ``low Reynolds number" is most commonly used to mean ``zero Reynolds number."}

The material derivative of the liquid fraction $\DDD{\varphi}{t}$ required on the right-hand side of the low Mach Eq.~\eqref{eq_divu} is obtained from the energy equation, which is written in terms of specific enthalpy $h$
\begin{equation}
\D{\left(\rho h\right)}{t}+\div{\left(\rho \u h\right)}= \rho \DDD{h}{t} = \div\left({\kappa\grad{T}}\right) + Q_{\rm src}.
\label{eq_enthalpy}
\end{equation}
Here, $Q_{\rm src}$ represents any heat source/sink term, such as a scanning laser beam. Note that the enthalpy equation is obtained by subtracting the kinetic energy equation from the conservation of energy equation by manipulating derivatives associated with velocity $\u$ and pressure $p$. For a diffuse interface formulation this is acceptable because velocity/kinetic energy and pressure are assumed to be continuous across the interface. For a sharp interface approach this leads to the loss of kinetic energy jump terms; see Sec.~\ref{sec_jump_conditions} for further discussion on jump conditions. 

Specific enthalpy $h$ of the PCM is defined in terms of its temperature $T$ as
\begin{align}
h = \begin{cases}
 \cps (T - T_r) ,&  T<\Tsol,\\ 
 \bar{C}(T - \Tsol) + \hsol + \varphi \frac{\displaystyle \rhol}{\displaystyle \rho}L,&\Tsol \le T \le \Tliq, \\ 
 \cpl(T-\Tliq)+\hliq,& T> \Tliq ,
\end{cases}
\label{eq_h_pcm}
\end{align}
and of the gas as
\begin{equation}
h = \cpg (T - T_r) .
\label{eq_h_gas}
\end{equation} 
In~Eq.~\eqref{eq_h_pcm}, $\hsol = \cps (\Tsol - T_r), \hliq = \bar{C}(\Tliq-\Tsol)+\hsol+L,  \text{ and  } \bar{C}=\frac{\cps+\cpl}{2}$. $\bar{C}$ is the specific heat of the mushy region, which is taken as an average of liquid and solid specific heats. Eqs.~(\ref{eq_h_pcm}) and (\ref{eq_h_gas}) imply that PCM and gas enthalpies are zero at $T = T_r$. \REVIEW{The numerical solution is not affected by this arbitrary choice of reference temperature $T_r$, and in the numerical simulations we set the melting/solidification temperature as the reference temperature $T_r = T_m$} \footnote{\REVIEW{Another reasonable choice is to set $T_r = 0$.}}. We use the mixture model~\cite{alexiades1993weak}  to express density and specific enthalpy in terms of liquid fraction in the mushy region 
\begin{align}
\rho &= \varphi \rhol + (1-\varphi) \rhos   \label{eq_rho_mixture}, \\
\rho h &= \varphi \rhol \hliq + (1-\varphi) \rhos \hsol.   \label{eq_h_mixture}
\end{align}
\REVIEW{Eqs.~\eqref{eq_rho_mixture} and~\eqref{eq_h_mixture} can also be derived from the general Eq.~\eqref{eq_material_properties} by substituting $H = 1$ (which holds true in the PCM region) and $\beta = \rho$ or $\rho h$.}

Substituting $h$ from Eq.~\eqref{eq_h_pcm} and $\rho$ from Eq.~\eqref{eq_rho_mixture} into Eq.~\eqref{eq_h_mixture}, we obtain a $\varphi$-$T$ relation for the mushy region 
\begin{equation}
\varphi = \frac{\displaystyle \rho}{\displaystyle \rhol}\frac{\displaystyle T-\Tsol}{\displaystyle \Tliq-\Tsol}.
\label{eq_varphi_mixture}
\end{equation}
Knowing $\varphi$ in terms of $T$ (Eq.~\eqref{eq_varphi_mixture}) allows us to invert $h$-$T$ relations. The  temperature in the PCM region
\begin{align}
T = \begin{cases}
 \frac{\displaystyle h}{\displaystyle \cps} + T_r, & h<\hsol,\\ 
  \Tsol + \frac{\displaystyle h-\hsol}{\displaystyle \hliq-\hsol}(\Tliq-\Tsol),&\hsol \le h \le \hliq, \\ 
 \Tliq + \frac{\displaystyle h-\hliq}{\displaystyle \cpl},& h> \hliq,
\end{cases}
\label{eq_T_pcm}
\end{align} 
and in the gas region
\begin{equation}
T =  \frac{\displaystyle h}{\displaystyle \cpg} + T_r
\label{eq_T_gas}
\end{equation}
can be written in terms of $h$. These $T$-$h$ relations are used in the Newton's iterations to solve the nonlinear~Eq.~\eqref{eq_enthalpy}. 
Similarly, substituting $\rho$ from Eq.~\eqref{eq_rho_mixture} into Eq.~\eqref{eq_h_mixture},  we get  a $\varphi$-$h$ relation 
\begin{align}
\varphi = \begin{cases}
 0,& h<\hsol,\\ 
  \frac{\displaystyle \rhos(\hsol-h)}{\displaystyle h(\rhol-\rhos)-\rhol \hliq + \rhos \hsol},&\hsol \le h \le \hliq, \\ 
 1,& h> \hliq.
\end{cases}
\label{eq_liquid_fraction}
\end{align}
Although arbitrary, $\varphi$ in the gas region is defined to be zero. 

Finally,  Eq.~\eqref{eq_liquid_fraction} allows us to define $\DDD{\varphi}{t}$ for the low Mach~Eq.~\eqref{eq_divu} as
\begin{align}
\DDD{\varphi}{t}=\begin{cases}
 0,& h<\hsol,\\ 
 \frac{\displaystyle -\rhos \rhol (\hsol-\hliq)}{\displaystyle (h(\rhol- \rhos)- \rhol \hliq + \rhos \hsol)^2}\displaystyle \DDD{h}{t},&\hsol \le h \le \hliq, \\ 
 0,& h> \hliq .
\end{cases}
\label{eq_dvarphi_dt}
\end{align}
The material derivative of $h$ in Eq.~\eqref{eq_dvarphi_dt} is obtained from the enthalpy Eq.~\eqref{eq_enthalpy} as \[\DDD{h}{t}=\frac{1}{\rho} \left( \div\left({\kappa\grad{T}}\right) + Q_{\rm src} \right).\] It is clear from Eq.~\eqref{eq_dvarphi_dt} that $\DDD{\varphi}{t} \ne 0$ only in the mushy region where $\hsol \le h \le \hliq$ and $\Tsol \le T \le \Tliq$. This results in a non-div-free velocity field in the mushy region, but div-free elsewhere. Therefore, in the absence of mushy regions, velocity is div-free. This can happen when a liquid phase has solidified completely or when a solid phase has melted completely. Our formulation, therefore, guarantees that there will be no change in the volume of the system in the absence of phase change. It can also be seen from Eq.~\eqref{eq_divu} that when the densities of the solid and liquid phases match, there is no induced flow and the velocity is div-free.   

The low Mach Eq.~\eqref{eq_divu} is solved in conjunction with the momentum equation 
\begin{align}
\D{\left(\rho \u\right)}{t}+\div{\left(\rho \u \otimes \u\right)} &= -\grad{p}+\div\left[{\mu\left(\grad{\u}+\grad{\u}^T\right)}\right] 
+\rho \g - A_d \u + \f_{\rm st},
\label{eq_momentum}
\end{align} 
to obtain the Eulerian velocity $\u(\x,t)$ and pressure $p(\x,t)$ in all three phases. Here, $\mu(\x,t)$ is the spatiotemporally varying viscosity that is defined using Eq.~\eqref{eq_material_properties}, $\g$ is the acceleration due to gravity, and $A_d= C_d\frac{\displaystyle {\varphi_{\rm{S}}}^2}{\displaystyle (1-\varphi_{\rm{S}})^3+\epsilon}$ is the Carman-Kozeny drag coefficient that is used to retard any flow in the solid domain,  $\varphi_{\rm{S}}=H(1-\varphi)$ is the solid fraction of the grid cell, and $\epsilon = 10^{-3}$ is a small number to avoid a division by zero and to control the strength of penalty parameter ($C_d/\epsilon$) in the solid region. \REVIEW{To retard fluid motion within the solid domain, the model parameter $C_d$ takes a large value. By comparing the magnitudes of the drag force and the first term on the left hand side of the momentum equation (i.e., equating intertial force to drag force), we obtain a sufficiently large value for $C_d=\rhos/\Delta t$. Here, $\Delta t$ denotes the time step size of the simulation.} $\f_{\rm st}$ is the surface tension force that acts on the liquid-gas interface. The next section details the numerical algorithm and the time stepping scheme. Due to the large density difference between the solid, liquid, and gas phases, special care is needed to avoid numerical instabilities. This is explained in the next section as well. Observe that the momentum equation is expressed using a diffuse interface formulation, where all quantities are assumed to vary continuously across the (three) phases. In addition, the Carman-Kozeny drag force strongly influences the pressure jump/gradient across the mushy region, which is similar to the Darcy-Brinkman equation for modeling flows in porous media~\cite{durlofsky1987analysis}. When a diffuse interface formulation is used for the momentum equation, $p$ and $\llbracket p \rrbracket$ will generally have numerical values that differ (perhaps by orders of magnitude) from those of a sharp interface formulation. This is discussed further in the context of the Stefan problem in the next section.        


 \subsection{Complete solution algorithm} \label{sec_solution_algo}

In this section we describe the time stepping algorithm used to solve the coupled mass, momentum and enthalpy equations described above. We assume that all quantities of interest, denoted $\theta$, have been computed or are known at time $t = t^n$.  To advance the solution to the next time level $n+1$, we employ $p$ number of fixed point iterations (with $k = 0, 1, \ldots, p -1$ denoting the iteration number) within a single time step to approximate $\theta^{n+1} =  \theta^{n+1, p-1}$. Within each fixed point iteration, we employ $q_\text{max}$ number of Newton's iterations (with $m = 0, 1, \ldots, q_\text{max}-1$ denoting the Newton's iteration number) to solve the nonlinear enthalpy equation. At the beginning of the time step, we initialize $\theta^{n+1, k = 0} = \theta^n$ for variables $\u, \rho, \phi$, and $H$.  The temperature variable is initialized similarly $T^{n+1, k = 0, m = 0}=T^n$. Hence, within a single time step (of size $\Delta t = t^{n+1} - t^n$), the Navier-Stokes system and the level-set and Heaviside advection equations are solved for $p$ times and the enthalpy equation is solved (possibly) for  $p \times q_\text{max}$ times. For all cases presented in this work, we use $p = 2$ fixed-point iterations and set $q_\text{max} = 5$ for the Newton solver, unless otherwise stated.  The governing equations are solved in the order described next.

\begin{enumerate}

\item  The level set function $\phi^n$ is first advected with the non-div-free velocity $\u$ to obtain $\phi^{n+1,k+1}$
\begin{equation}
\frac{\phi^{n+1,k+1}-\phi^{n}}{\Delta t} + \left( \div [\phi \u] \right)^{n+1,k} = (\phi\div{\u})^{n+1,k}. \label{eq_phi_advect}
\end{equation}
Under (linear) advection, $\phi$ does not maintain its signed distance property. A reinitialization procedure suggested by Sussman et al.~\cite{Sussman1994} is used to restore the signed distance property of $\phi$. Implementation details about the level set method, and its reinitialization can be found in our prior work \cite{nangia2019robust}. 

\item A smooth Heaviside function $H$ is used to track the gas-PCM interface. $H$ takes a value of 0 in the gas region, 1 in the solid-liquid PCM region, and transitions smoothly from 0 to 1 around the interface with a prescribed width of $n_\text{cells} = 2 $ grid cells (of size $\Delta$)  on either side of the gas-PCM interface 
\begin{align}
H &= \begin{cases}
 0,& \phi(\x)<-n_\text{cells} \, \Delta,\\ 
 1-\frac{1}{2}\left[1+\frac{1}{n_\text{cells} \, \Delta} \; \phi(\x) +  
     \frac{1}{\pi}\text{sin}\left(\frac{\pi}{n_\text{cells} \, \Delta} \; \phi(\x)\right)\right],& |\phi(\x)| \le n_\text{cells} \, \Delta,\\ 
 1,&\text{otherwise}.
\end{cases}
\label{eq_smooth_H}
\end{align}
Although $H^{n+1,k+1}$ can be defined directly in terms of $\phi^{n+1,k+1}$, we instead advect $H^n$ (defined in terms of $\phi^n$ using~Eq.~\eqref{eq_smooth_H}) to approximate  $H^{n+1,k+1}$
\begin{equation}
\frac{H^{n+1,k+1}-H^{n}}{\Delta t} + \left(\div [H \u]\right)^{n+1,k} = (H\div{\u})^{n+1,k}. \label{eq_H_advect}
\end{equation}
This is done to obtain the advective flux of Heaviside $H\u$, which could be used to advect additional scalar variables of a more involved problem. At the end of the time step $H^{n+1}$ is synchronized with $\phi^{n+1}$ using Eq.~\eqref{eq_smooth_H}. A third-order accurate cubic upwind interpolation (CUI) scheme \cite{nangia2019robust} is used for advecting $\phi$ and $H$ in Eqs.~(\ref{eq_phi_advect}) and (\ref{eq_H_advect}). CUI satisfies both the convection-boundedness criterion (CBC) (see chapter 12 of \cite{moukalled2016finite} for a discussion on high-resolution schemes) as well as the total variation diminishing (TVD) property. Both properties are essential to bound $H$ ($0 \le H \le 1$) during its advection.    



\item In practical applications, the density contrast between PCM and gas is usually very large. The metal to gas density ratio, for example, is $\rhos/\rhog \sim 10^4$. It is important to ensure numerical stability of the scheme when advecting energy/enthalpy and momentum in the domain with very high density ratios. Our recent work proposed an efficient approach for maintaining the stability of isothermal flows (no phase change) with a high density ratio. It involves solving an additional mass balance equation and computing the mass flux $\V{m_\rho} = \rho \u$. The same mass flux $\V{m_\rho}$ is  used in the convective operator of the momentum equation, i.e., $\div (\rho \u \otimes \u)$ is discretized as $\div (\V{m_\rho} \otimes \u)$ in the momentum equation. The same idea is applied to advect enthalpy $h$ as well, i.e., $\div (\rho \u h) = \div (\V{m_\rho} h)$. 

The discrete mass balance equation reads as
\begin{equation}
\frac{\breve{\rho}^{n+1,k+1}-\rho^{n}}{\Delta t} + (\div  \V{m_\rho})^{n+1,k} = 0, \label{eq_mass_advect}
\end{equation}
which is solved to obtain $\breve{\rho}^{n+1,k+1}$ and the discrete mass flux $\V{m_\rho}$. We use a second-order accurate explicit Runge-Kutta scheme (mid-point rule) for time integrating Eq.~\eqref{eq_mass_advect}. CUI is used as a limiter to ensure that $\rho$ remains bounded during advection.  In Eq.~\eqref{eq_mass_advect} $\rho^n$ is defined through EOS. In other words, approximation to the new density $\breve{\rho}^{n+1,k+1}$ is only temporarily used within a time step, after which it is synchronized with the EOS. The synchronization step ensures that (\textbf{i}) density does not deviate from the EOS; and (\textbf{ii})  gas-PCM interface remains sharp. The latter is due to the use of reinitialized level set function in the EOS. 

\item Using the discrete approximation for the new density $\breve{\rho}$ and mass flux $\V{m_\rho}$, the nonlinear enthalpy equation is solved to update enthalpy $h$, temperature $T$, and liquid fraction $\varphi$. $\kappa$ and $C$ are defined as functions of $\varphi$, which is in turn a function of $h$. As a result, the enthalpy equation is highly nonlinear. 

The discrete enthalpy equation reads as
\begin{equation}
\frac{\breve{\rho}^{n+1,k+1} h^{n+1,k+1}-\rho^n h^{n}}{\Delta t} + (\div \V{m_\rho} h)^{n+1,k} = (\div\kappa\grad T)^{n+1,k+1} + Q_{\rm src}.
\label{eq_energy_discretized}
\end{equation}
In the examples considered in this work, there are no additional heat source/sink terms, i.e., $Q_\text{src} = 0$ in Eq.~\eqref{eq_energy_discretized}. This term could be treated implicitly or explicitly depending upon the numerical stiffness and/or its complexity. We omit the $Q_\text{src}$ term in the remainder of the algorithm.

\begin{enumerate}

\item The nonlinear enthalpy equation is solved using Newton's iteration to obtain $h^{n+1,k+1}$. Specifically, $h$ is linearized using Taylor's expansion as
\begin{equation}
h^{n+1,k+1,m+1} = h^{n+1,k+1,m} + \D{h}{T}\bigg|^{n+1,k+1,m}(T^{n+1,k+1,m+1}-T^{n+1,k+1,m}),
\label{eq_enthalpy_linearization}
\end{equation}
in which $m$ is the inner (Newton) iteration level. Substituting the above equation into Eq.~\eqref{eq_energy_discretized}, we get
\begin{align}
\frac{\tilde{\rho}^{n+1,k+1}\left( h^{n+1,k+1,m} + \D{h}{T}\bigg|^{n+1,k+1,m}\left(T^{n+1,k+1,m+1}-T^{n+1,k+1,m} \right)\right)-\rho^n h^{n}}{\Delta t} + \div{(\V{m_\rho} h)}^{n+1,k} \nonumber \\= (\div\kappa\grad T)^{n+1,k+1,m+1}
\label{eq_temp_discretized}
\end{align}
Eq.~\eqref{eq_temp_discretized} is solved to obtain $T^{n+1,k+1,m+1}$. The $h$-$T$ relations written in Eq.~\eqref{eq_h_pcm} allows an analytical evaluation of $\D{h}{T}$. 
\REVIEW{
Specifically, in the PCM domain, the derivative $\D{h}{T}$ is given by  
\begin{align}
\D{h}{T}\Bigg|_\text{PCM} = \begin{cases}
 \cps,&  T<\Tsol,\\ 
 \bar{C}+L/(\Tliq-\Tsol),&\Tsol \le T \le \Tliq, \\ 
 \cpl,& T> \Tliq ,
\end{cases}
\label{eq_dhdt_pcm}
\end{align}
and in the gas $\D{h}{T}\Big|_\text{gas} = \cpg$. Note that the specific enthalpy of the PCM is defined to be a $\mathcal{C}^0$ piecewise-continuous function\footnote{\REVIEW{Strictly speaking, the specific enthalpy $h$ of a pure PCM cannot be a $\mathcal{C}^0$ continuous function. This is because a large amount of latent heat is released/absorbed at its solidification/melting temperature $T_m$ and $h$ jumps at $T_m$. In enthalpy methods, this condition is relaxed and the latent heat is assumed to be released over a temperature interval $\Delta T = \Tliq - \Tsol$.}} of $T$, and its derivative (with respect to $T$) jumps at $\Tsol$ and $\Tliq$. 
The PCM and gas regions are distinguished by the Heaviside contour $H = 0.5$ (or alternatively by the $\phi = 0$ contour). Therefore, 
$\D{h}{T}$  in the entire domain is defined as: 
\begin{align}
\D{h}{T} = \begin{cases}
 \D{h}{T}\Big|_\text{PCM},&  H\ge0.5,\\  \\
 \D{h}{T}\Big|_\text{gas},& \text{otherwise}.
\end{cases}
\label{eq_dhdt_domain}
\end{align}
$\D{h}{T}$ defined in Eq.~\eqref{eq_dhdt_domain} can be made ``more smooth" by defining it as $\D{h}{t} = H \D{h}{T}\Big|_\text{PCM} + (1 - H) \D{h}{T}\Big|_\text{gas}$. In our numerical experiments we did not observe any 
major improvement in the Newton solver (in term of its convergence rate) using the ``smoother" version of  $\D{h}{t}$. Hence, we make use of Eq.~\eqref{eq_dhdt_domain} in the code. 
}
The linear system of Eq.~\eqref{eq_temp_discretized} is solved using a geometric multigrid preconditioned FGMRES solver with a relative tolerance of $10^{-9}$.

\item Update enthalpy $h^{n+1,k+1,m+1}$ using the Taylor series expansion (Eq.~\eqref{eq_enthalpy_linearization}) and $T^{n+1,k+1,m+1}$.

\item Update $T^{n+1,k+1,m+1}$ and $\varphi^{n+1,k+1.m+1}$ using $h^{n+1,k+1,m+1}$ and analytical $T$-$h$ and $\varphi$-$h$ relations written in the Sec.~\ref{sec_math_formulation}, respectively. 

\item Update thermophysical properties ($\kappa, C$, and $\mu$) using $\varphi^{n+1,k+1, m+1}$.  \REVIEW{In spite of the fact that the model does not need specific heat $C$ directly, we update it for consistency reasons. The updated $C$ values can be used for post-processing or to model additional physics, for example.}

\item Compute the relative change in liquid fraction 
\begin{equation}
\epsilon = \frac{||\varphi^{n+1,k+1,m+1}- \varphi^{n+1,k+1,m}||_2}{||1+ \varphi^{n+1,k+1,m}||_2}
\end{equation}
The Newton solver is deemed to be converged if $\epsilon \le 10^{-8}$ or if $m+1 = q_\text{max} = 5$ iterations have completed.  
\end{enumerate}

\item Finally we solve the momentum and low Mach equations together 
\begin{alignat}{2}
& \frac{\breve{\rho}^{n+1,k+1} \u^{n+1,k+1}-\rho^n \u^{n}}{\Delta t} + (\div [\V{m_\rho} \otimes \u])^{n+1,k}  = -{\grad p}^{n+\frac{1}{2},k+1}+(\div\left[\mu\left(\grad \u+\grad \u^T\right)\right])^{n+\frac{1}{2},k+1}   \nonumber  \\
&\hspace{22em} - A_d^{n+1,k+1} \u^{n+1,k+1} + {\V{f}_{\rm st}}^{n+\frac{1}{2},k+1}, \label{eq_momentum_discretized} \\
& \div{\u} = \begin{cases} 
0, & H < 0.5 \; (\text {i.e., in the gas phase}), \\
0,& h<\hsol,\\ 
-\frac{ \rhos \rhol}{\rho^2}(\rhol-\rhos)H \frac{\displaystyle (\hliq-\hsol)}{\displaystyle \left(h(\rhol- \rhos)- \rhol \hliq +\rhos \hsol \right)^2}\displaystyle \left(\div {\kappa\grad{T}}  \right),&\hsol \le h \le \hliq, \\ 
0,& h>\hliq.\\ 
\end{cases} \label{eq_lowmach_discretized}
\end{alignat}
to update velocity $\u^{n+1,k+1}$ and pressure $p^{n+\frac{1}{2},k+1}$. In Eq.~\eqref{eq_momentum_discretized} we use the same discrete density $\breve{\rho}^{n+1,k+1}$ and mass flux $\V{m_\rho}$ that we obtained from solving Eq.~\eqref{eq_mass_advect}. This maintains consistency between mass and momentum transport for high-density ratio flows. The Carman-Kozeny drag coefficient $A_d$ employs an updated value of $\varphi^{n+1,k+1}$ obtained from solving the enthalpy Eq.~\eqref{eq_energy_discretized}. The surface tension force $\V{f}_{\rm st}$ acting on the liquid-gas interface is modeled using the continuous surface tension formulation~\cite{brackbill1992continuum,saldi2012marangoni,francois2006balanced}. The continuous surface tension force reads as 
\begin{align}
{\V{f}_{\rm st}} =  \varphi \frac{2 \breve{\rho}}{\rhol +\rhog} \left(\sigma \mathcal{C} \grad {\widetilde{B}} \right), \label{eq_surface_tens}
\end{align}
in which $\sigma$ is the uniform liquid-gas surface tension coefficient and $\mathcal{C}(\phi)$ is the curvature of the interface computed from the level set function $\mathcal{C}^{n+\frac{1}{2},k+1} = - \div \left( \frac{\grad \phi}{|| \grad \phi ||} \right) $.  In Eq.~\eqref{eq_surface_tens}, $\widetilde{B}(\phi)$ represents a mollified Heaviside function that ensures the surface tension force only acts near the PCM-gas region. In addition, the multiplier $\varphi$ limits the influence of surface tension to liquid and gas. We use the mid-point value of $\phi^{n+\frac{1}{2},k+1} = \frac{1}{2}\left(\phi^{n+1,k+1} + \phi^n \right)$ in computing $\mathcal{C}$ and $\widetilde{B}$. The right-hand side of the discrete low Mach Eq.~\eqref{eq_lowmach_discretized} is evaluated by using the most updated values of $H^{n+1,k+1}, h^{n+1,k+1}, T^{n+1,k+1}$, and $\rho^{n+1,k+1}$. The linear system of Eqs.~(\ref{eq_momentum_discretized}) and (\ref{eq_lowmach_discretized}) is solved monolithically for the coupled $\u$-$p$ system using an FGMRES solver with a relative tolerance of $10^{-9}$. The FGMRES solver employs a projection method-based preconditioner as explained in Thirumalaisamy et al.~\cite{thirumalaisamy2023pre}.  


\end{enumerate}

Each case presented in this work uses a uniform time step size $\Delta t$, and the CFL number does not exceed 0.5.  All of the spatial derivatives appearing in Eqs.~(\ref{eq_phi_advect})-(\ref{eq_lowmach_discretized}) are approximated using second-order accurate finite differences. \REVIEW{In Appendix~\ref{sec_spatial_discretization} we briefly describe the spatial discretization framework employed in this work}. More details on the discretization technique and variable coefficient (density and viscosity) flow solver are provided in our prior works~\cite{nangia2019robust,Nangia2019WSI}.  

\section{Software implementation}    
The numerical algorithm detailed above is implemented within the IBAMR library~\cite{IBAMR-web-page}. 
IBAMR is an open-source C++
software enabling simulation of CFD and fluid-structure interaction problems on block-structured Cartesian grids. 
The code is hosted on GitHub at \REVIEW{\url{https://github.com/IBAMR/IBAMR/pull/1627}}.
IBAMR relies on SAMRAI \cite{HornungKohn02,samrai-web-page} for Cartesian grid 
management. Solver support in IBAMR is provided by the 
PETSc library~\cite{petsc-user-ref, petsc-web-page}. CFD results are post-processed using a combination of in-house 
MATLAB and Python scripts, as well as using the open-source VisIt visualization software \cite{childs2012visit}.

\section{Results and discussion}  \label{sec_results}
\subsection{Validation of the low Mach enthalpy method with analytical solutions}  \label{sec_metal_melting}
 
We validate the novel low Mach EM by examining the Stefan problem (solidification of a liquid PCM) discussed in Sec.~\ref{sec_analytical_stefan}, with three different density ratios $R_\rho = \rhos/\rhol$ that lead to: (\textbf{1}) no flow ($R_\rho= 1$); (\textbf{2}) flow with volume expansion ($R_\rho < 1$); and (\textbf{3}) flow with volume shrinkage ($R_\rho > 1$). The numerical model consists of a quasi one-dimensional computational domain $\Omega \in [0,1] \times [0, 0.05]$ with $N_x \times N_y = 1280 \times 64$ grid cells. The domain is periodic in the $y$-direction. Initially, liquid occupies the entire domain at $T_i = 973.6$ K. The left boundary ($x=0$) is set to $T_o = 298.6$ K and the right boundary ($x=1$) is adiabatic (homogeneous Neumann). The flow solver uses zero-velocity and zero-pressure/outflow boundary conditions at the left and right ends, respectively. PCM's thermophysical properties are largely aluminum-based, and are listed in Table~\ref{tab_aluminum_properties}. In this case, both fluid and solid viscosities are set to zero. For simplicity, we take the bulk phase change temperature $T_r$ equal to the solidification temperature $T_m$, so $L^{\rm eff} = L$. 
\REVIEW{The temperature interval $\Delta T$ between solidus and liquidus and the grid size are selected based on convergence studies presented in Secs.~\ref{sec_stefan_deltaT_convergence_study} and~\ref{sec_stefan_grid_convergence_study}, respectively}.  
\begin{table}
\centering
\caption{Thermophysical properties used to simulate the Stefan problem}
\label{tab_aluminum_properties}
\begin{tabular}{ll}
Property & Value \\
\midrule
Thermal conductivity of solid   $\ks$     &   211  W/m.K  \\
Thermal conductivity of liquid  $\kl$                 & 91  W/m.K    \\
Specific heat of solid  $\cps$                 & 910 J/kg.K      \\
Specific heat of liquid  $\cpl$                  &  1042.4  J/kg.K   \\
Solidification temperature  $T_m$                &  933.6 K     \\
Bulk phase change temperature $T_r$                   & 933.6 K \\
Liquidus temperature $\Tliq$                       & 938.6 K \\
Solidus temperature $\Tsol$                       & 928.6 K  \\
Latent heat  $L$                &  383840  J/kg    \\
\bottomrule
 \end{tabular}
\end{table}
 
\subsubsection{The no volume change case}  \label{sec_stefan_novolchange}

\begin{figure*}
\centering
\includegraphics[width=1.0\linewidth]{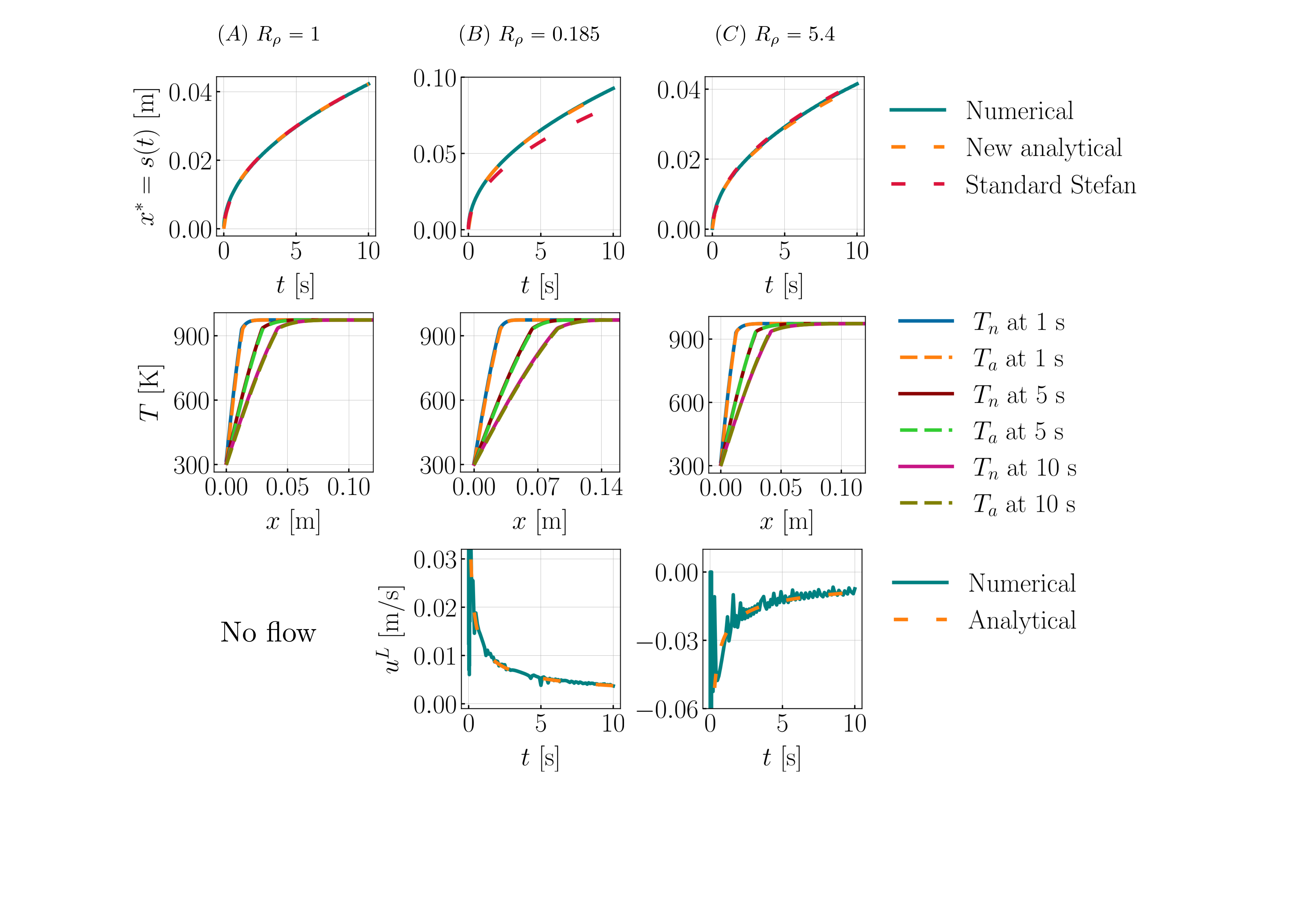}
\caption{Comparison of CFD and analytical solutions to the Stefan problem at various density ratios $R_\rho = \rhos/\rhol$. (A) Comparison of the solid-liquid interface position and temperature distribution in the domain when the liquid and solid densities are the same. In this case, there is no fluid flow. (B) and (C), respectively, compare CFD and analytical solutions (interface position and temperature and [uniform] velocity distributions) for the expansion ($R_\rho =0.185)$ and shrinkage ($R_\rho=5.4$) cases.}
\label{fig_stefan_problem}
\end{figure*}

Fig.~\ref{fig_stefan_problem}(A)  compares CFD results\footnote{\REVIEW{This benchmark test is provided in IBAMR GitHub  within the directory \texttt{examples/phase\_change/ex0}.}} for the interface position $x^* = s(t)$ and temperature distribution in the solid and liquid phases  against the analytical solutions for $R_\rho=1$ case. We take solid and liquid densities to be the same $\rhos = \rhol =  2475$  kg/m$^3$. Table~\ref{tab_aluminum_properties} lists the rest of the thermophysical properties. Simulation is run until $t = 10$ s with a uniform time step size of $\Delta t = 10^{-3}$ s. The numerical solid-liquid interface is defined by an iso-contour value of 0.5 of the liquid fraction $\varphi$. It is evident from the figure that the interface position and temperature profiles match the analytical solution very well at different times. The analytical solution derived in this work reduces to the solution of the standard Stefan problem when $R_\rho$ equals 1. The top row of Fig.~\ref{fig_stefan_problem} shows both the new and standard Stefan problem solutions for $x^*$; the latter solution is from the Hahn and \"{O}zi\c{s}ik textbook \cite{hahn2012heat}.


\subsubsection{The expansion case} \label{sec_stefan_expansion}

For this case, the liquid and solid densities are assumed to be $\rhol = 2700$ and $\rhos = 500$  kg/m$^3$, respectively. Other thermophysical properties of aluminum-based PCM can be found in Table~\ref{tab_aluminum_properties}.  A uniform time step size of $\Delta t = 10^{-4}$ s is used throughout the simulation to maintain the CFL number below 0.5. Analytical and CFD solutions\footnote{\REVIEW{This benchmark test is provided in IBAMR GitHub within the directory \texttt{examples/phase\_change/ex1}.}} are compared in Fig.~\ref{fig_stefan_problem}(B). There is excellent agreement between the two. As can be seen, the standard Stefan solution underpredicts the interface position. This is because it does not take into account the additional flow that is generated in the direction of interface propagation. In addition, the temperature and liquid velocity profiles agree well with the new analytical model.  At $t = 0^+$ the interface velocity $u^*= \d s/ \d t \; \propto \; 1/\sqrt{t} \rightarrow \infty$. As  fluid velocity is proportional to interface speed (see Eq.~\eqref{eqn_vel_jump}), the CFD simulation produces large $\ul$ values at the beginning\footnote{At $t = 0$, the liquid is taken to be quiescent. $\ul$ starts with a zero value but  jumps to a large value immediately for CFD velocity profiles.}. The pressure profiles from CFD and analytical methods for this case are compared in Sec.~\ref{sec_stefan_pr_jump}.   

\subsubsection{The shrinkage case} \label{sec_stefan_shrinkage}

In order to simulate shrinkage, liquid and solid densities are assumed to be $\rhol = 500$ and $\rhos = 2700$  kg/m$^3$, respectively.  All other simulation and thermophysical parameters are kept the same as in the expansion case. Results\footnote{\REVIEW{This benchmark test is provided in IBAMR GitHub within the directory \texttt{examples/phase\_change/ex1}.}} are shown in Fig.~\ref{fig_stefan_problem}(C). Liquid-solid interface location matches the analytical solution very well.  Temperature and velocity profiles are also in good agreement with the analytical solution. $R_\rho > 1$ results in fluid flow opposite to the interface propagation, since fluid shrinks as it solidifies. Solidification rate is (slightly) reduced as additional hot fluid is pulled towards the solidifying front. Both the new analytical solution and standard Stefan  solution (without a density jump) predict an interface position that is qualitatively similar. Volume shrinkage during solidification may seem insignificant based on this analysis. This argument is refuted in Sec.~\ref{sec_metal_solidification}, in which we present a modeling study that highlights the importance of volume shrinkage in causing pipe defects during metal casting.

\REVIEW{ 
\subsubsection{$\Delta T$  convergence study for the Stefan problem with volume change} \label{sec_stefan_deltaT_convergence_study}
}

The thickness of the mushy zone for the enthalpy method depends on the temperature interval $\Delta T = \Tliq - \Tsol$ around the phase change temperature $T_m$. The numerical solution is expected to approach the analytical one as $\Delta T \rightarrow 0$. In practice $\Delta T$ is kept finite so that the latent heat can be absorbed or released within the grid-resolved mushy region. While simulating, if $\Delta T$ is set too small (but finite), the mushy region becomes very narrow and falls in the sub-grid region. Numerical oscillations are produced by intermittent appearances and disappearances of the mushy region during simulation. In order to select grid size and temperature interval for the EM, a convergence study is necessary.

\begin{figure}
\centering
\includegraphics[width=0.4\linewidth]{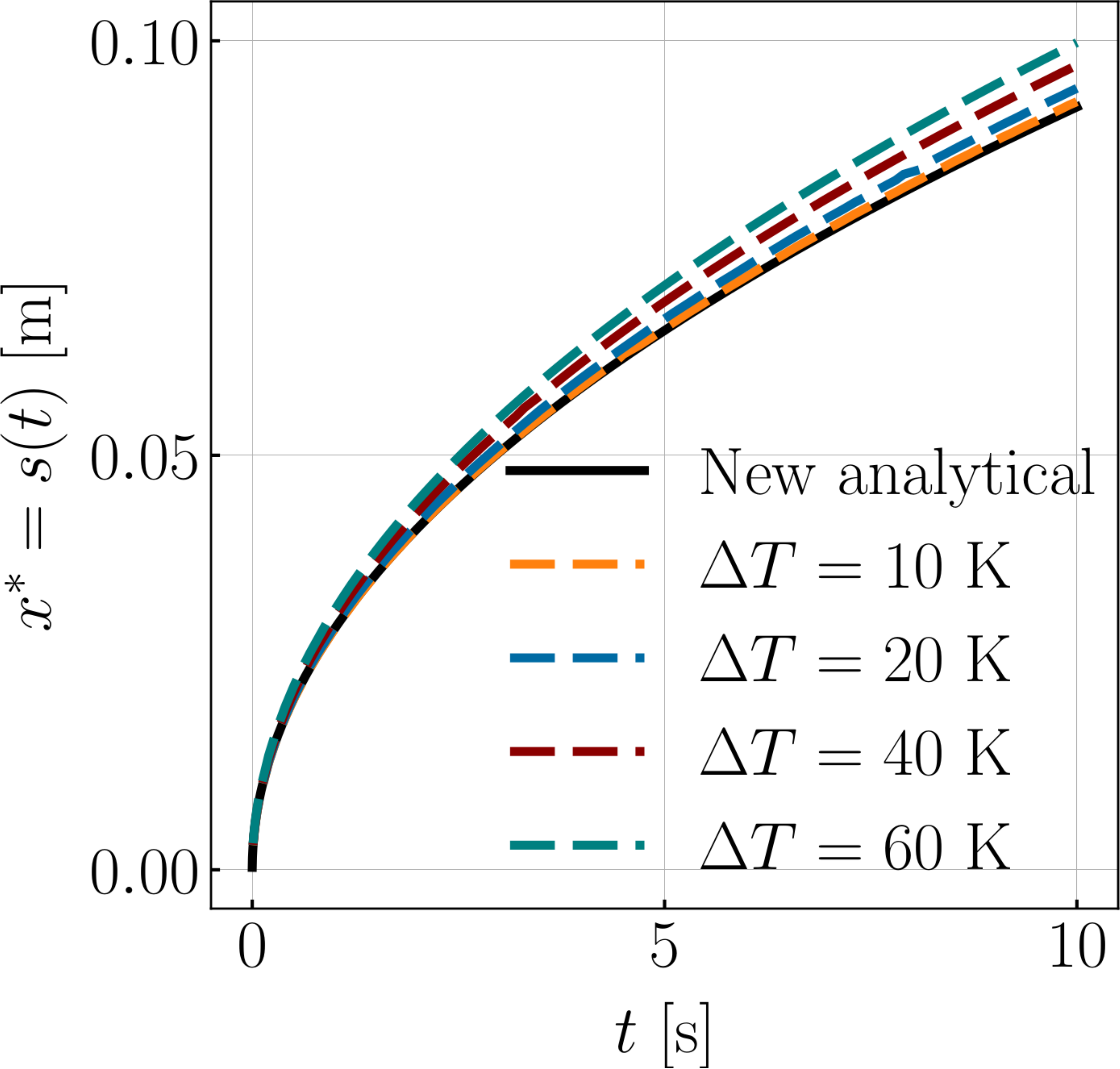}
\caption{Effect of temperature interval $\Delta T = \Tliq - \Tsol$ on the numerical solution of the Stefan problem with volume expansion ($R_\rho = 0.185$). The grid size considered is $N_x \times N_y = 1280 \times 64$. A uniform time step size of $\Delta t=10^{-4}$ is used for all the temperature intervals.}
\label{fig_stefan_deltaT_effect}
\end{figure}

\begin{figure}
\centering
\includegraphics[width=0.4\linewidth]{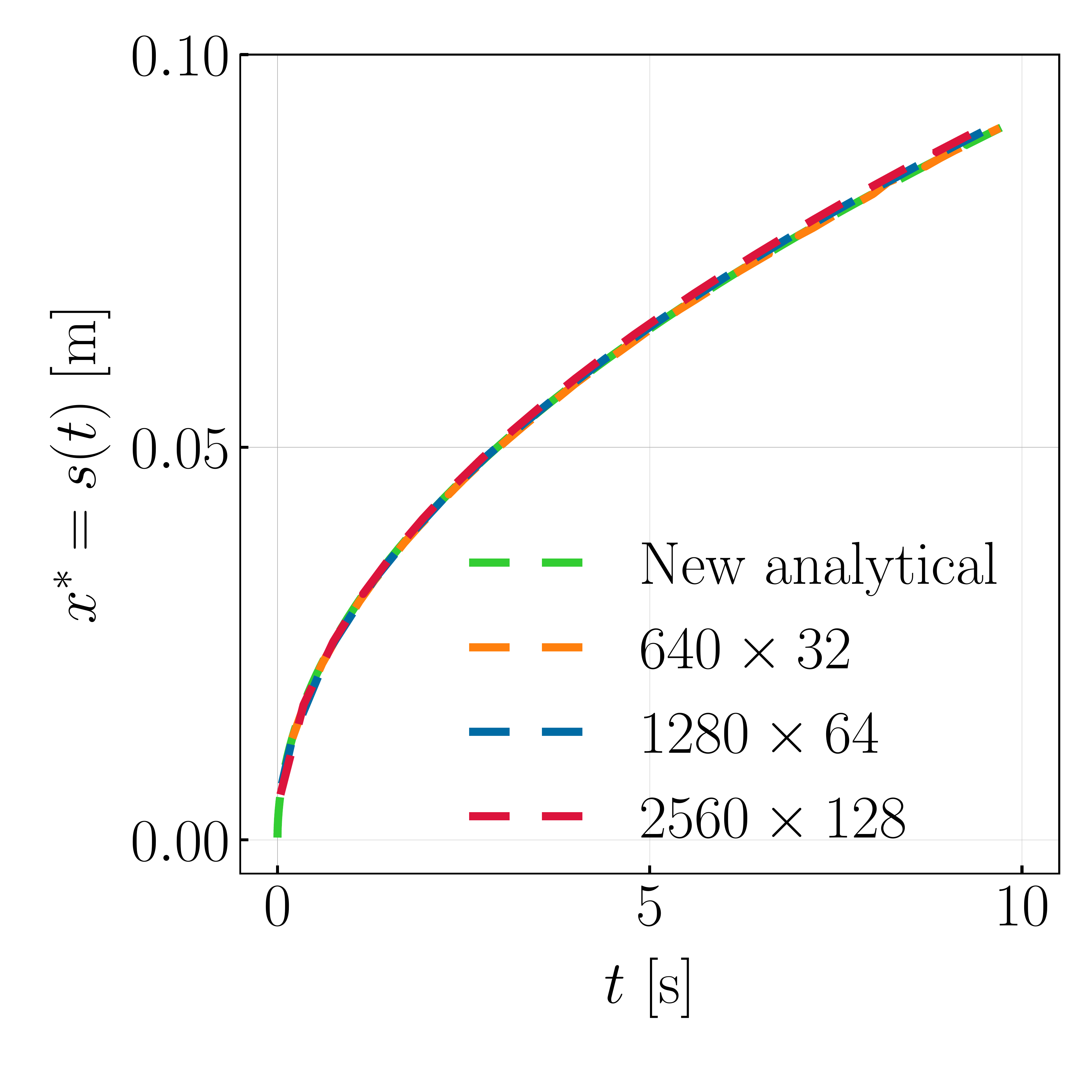}
\caption{Grid convergence study for the Stefan problem with volume expansion ($R_\rho=0.185$).  $\Delta T=10$ K is used for all grids. Uniform time step sizes used for the coarse, medium, and fine grids are $\Delta t=10^{-3}, 10^{-4}$, and $5 \times 10^{-5}$ s, respectively.}
\label{fig_stefan_grid_effect}
\end{figure}

For the Stefan problem simulated in this section, we consider $\Delta T = \{10, 20, 40, 60\}$ K. A convergence study is performed for the expansion problem $R_\rho=0.185$. A fixed grid size of $Nx \times Ny = 1280 \times 64$ is chosen for the study. Results for the interface position $x^* = s(t)$ at various temperature intervals are shown in Fig.~\ref{fig_stefan_deltaT_effect}. As expected, a smaller temperature interval leads to a more accurate solution for the EM. Additionally, we also considered $\Delta T$ = 2.5 and 5 K; the numerical solutions  either did not change appreciably or exhibited minor oscillations at coarse grid resolutions for these values of $\Delta T$ (data not presented). Therefore, we use $\Delta T = 10 \text{ K}$ in the numerical simulations for a PCM that is largely aluminum-based, unless otherwise stated. We also perform a grid convergence study using three grids: coarse, medium and fine of size $N_x  \times N_y =640 \times 32,1280 \times 64, 2560 \times 128$, respectively. $\Delta T = 10$ K is used for three grids. As observed in Fig.~\ref{fig_stefan_grid_effect}, the analytical and numerical solutions agree reasonably well. Consequently, we use medium grid to simulate the Stefan problems of this section. 

\begin{figure}
\centering
\includegraphics[width=0.8\linewidth]{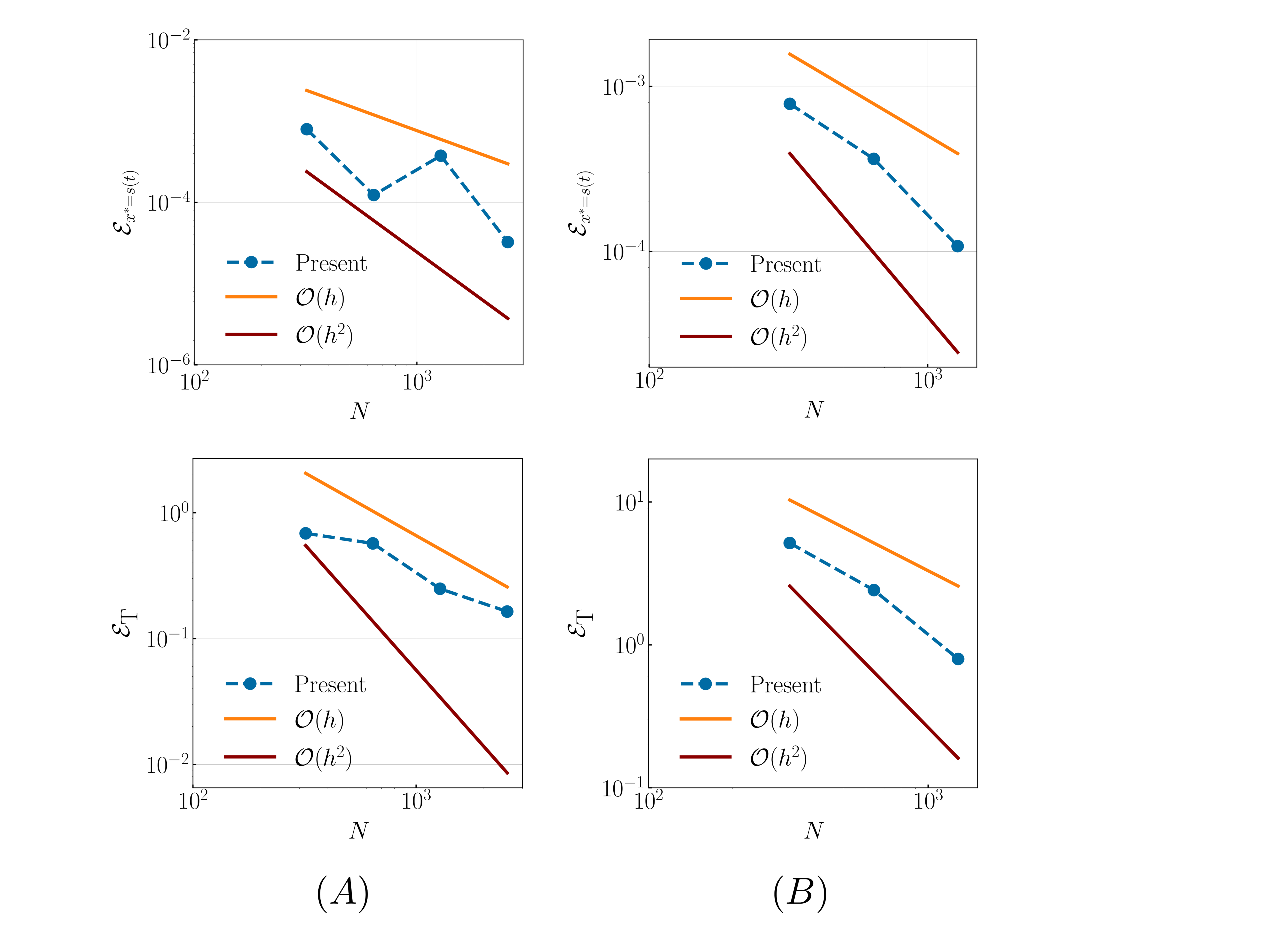}
\caption{\REVIEW{Convergence rates of the low Mach enthalpy method considering the Stefan problem with matched densities of solid and liquid phases ($\rhol=\rhos$).  The $\mathcal{L}^2$ error for a quantity $\psi$ is defined to be the root mean squared error (RMSE) of the vector $||\mathcal{E}_\psi ||_{\rm RMSE} =||\psi_\text{reference}-\psi_\text{numerical}||_2/\sqrt{\mathcal{N}}$, in which $\mathcal{N}$ denotes the size of the vector $\mathcal{E}_\psi$. Here, $\Psi$ represents the interface position $x^* = s(t)$ and temperature $T(x,t)$ in the domain. The reference solutions are (A) the analytical solutions and (B) the finest grid ($ N_x \times N_y = 2560\times 128$) numerical solutions. Errors are presented as a function of mesh resolution for the interface position $x^*=s(t)$ for the entire simulation period ($0 \le  t \le 10$ s) and for temperature $T(x,t)$ in the entire domain ($0 \le x \le l$) at $t=5$ s. The temperature interval between liquidus and solidus is $\Delta T=10$ K.}}
\label{fig_ex0_convergence_rate}
\end{figure}

\begin{figure}
\centering
\includegraphics[width=0.8\linewidth]{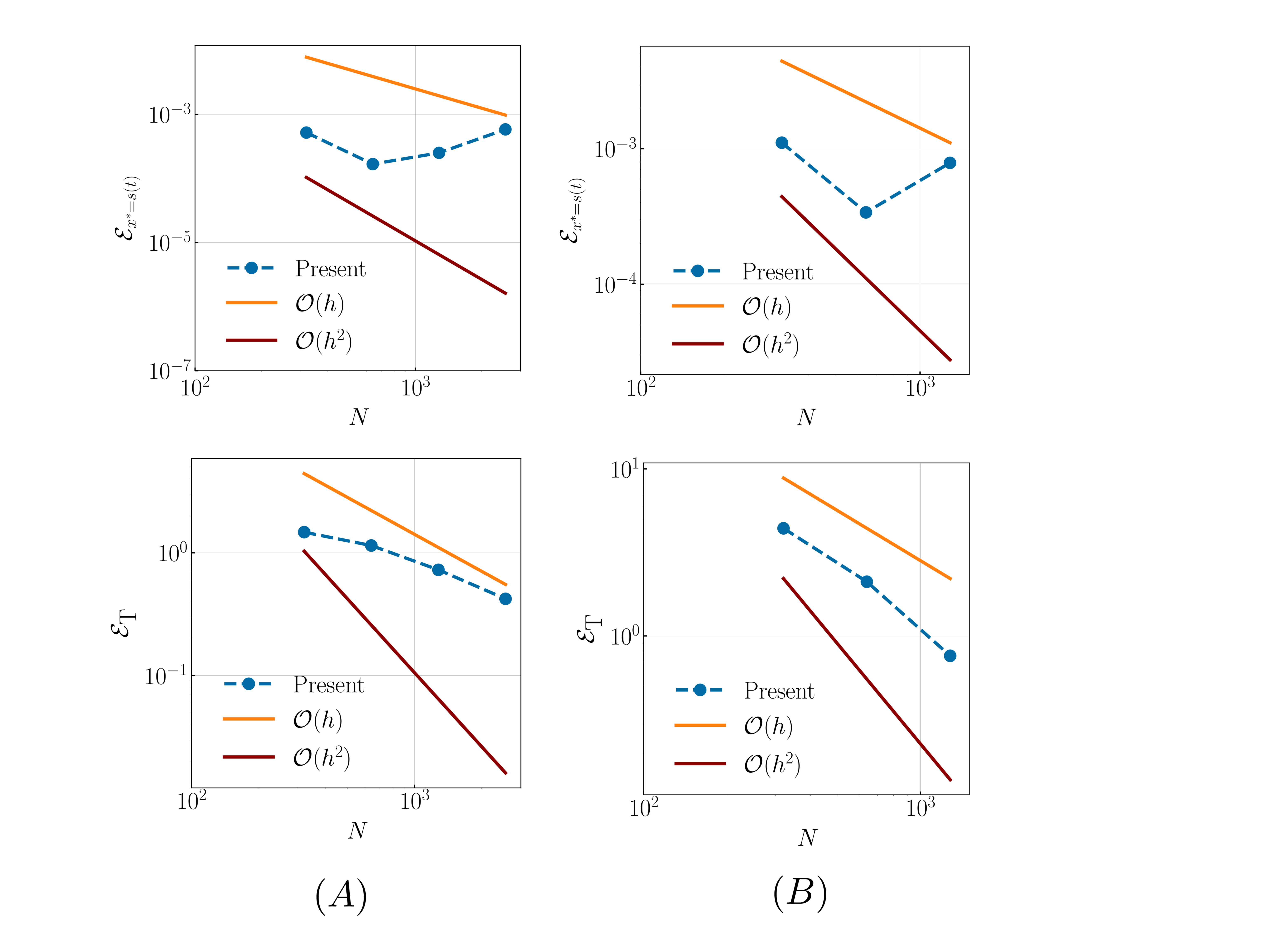}
\caption{\REVIEW{Convergence rates of the low Mach enthalpy method considering the Stefan problem with the solid phase density lower than the liquid phase ($R_\rho<1$).  The $\mathcal{L}^2$ error for a quantity $\psi$ is defined to be the root mean squared error (RMSE) of the vector $||\mathcal{E}_\psi ||_{\rm RMSE} =||\psi_\text{reference}-\psi_\text{numerical}||_2/\sqrt{\mathcal{N}}$, in which $\mathcal{N}$ denotes the size of the vector $\mathcal{E}_\psi$. Here, $\Psi$ represents the interface position $x^* = s(t)$ and temperature $T(x,t)$ in the domain. The reference solutions are (A) the analytical and (B) the finest grid ($ N_x \times N_y = 2560\times 128$) numerical solutions. Errors are presented as a function of mesh resolution for the interface position $x^*=s(t)$ for the entire simulation period ($0 \le  t \le 10$ s) and for temperature $T(x,t)$ in the entire domain ($0 \le x \le l$) at $t=5$ s. The temperature interval between liquidus and solidus is $\Delta T=10$ K.}}
\label{fig_ex1_expansion_convergence_rate}
\end{figure}

\begin{figure}
\centering
\includegraphics[width=0.8\linewidth]{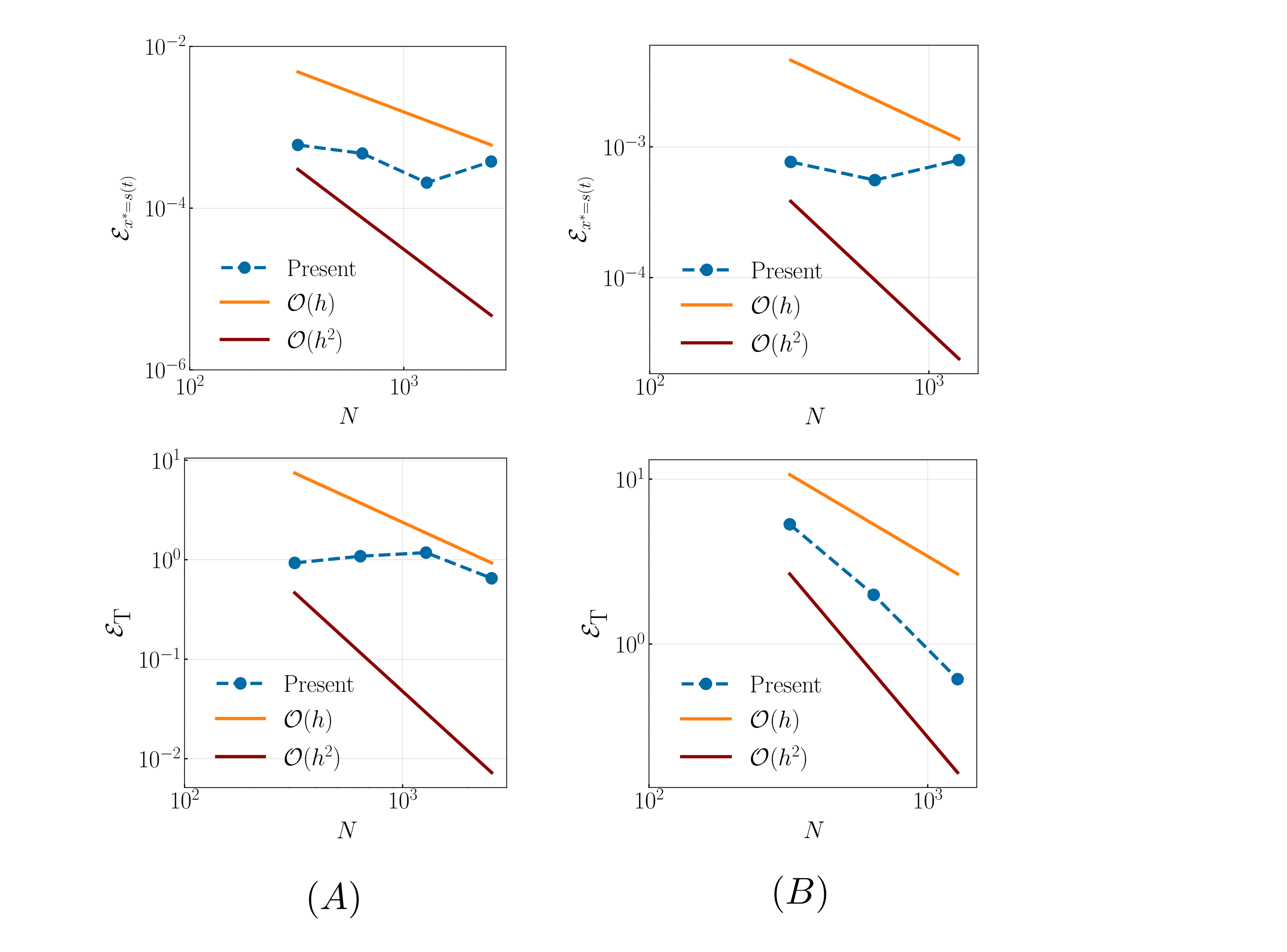}
\caption{\REVIEW{Convergence rates of the low Mach enthalpy method considering the Stefan problem with the solid phase density higher than the liquid phase ($R_\rho > 1$).  The $\mathcal{L}^2$ error for a quantity $\psi$ is defined to be the root mean squared error (RMSE) of the vector $||\mathcal{E}_\psi ||_{\rm RMSE} =||\psi_\text{reference}-\psi_\text{numerical}||_2/\sqrt{\mathcal{N}}$, in which $\mathcal{N}$ denotes the size of the vector $\mathcal{E}_\psi$. Here, $\Psi$ represents the interface position $x^* = s(t)$ and temperature $T(x,t)$ in the domain. The reference solutions are (A) the analytical and (B) the finest grid ($ N_x \times N_y = 2560\times 128$) numerical solutions. Errors are presented as a function of mesh resolution for the interface position $x^*=s(t)$ for the entire simulation period ($0 \le  t \le 10$ s) and for temperature $T(x,t)$ in the entire domain ($0 \le x \le l$) at $t=5$ s. The temperature interval between liquidus and solidus is $\Delta T=10$ K.}}
\label{fig_ex1_shrinkage_convergence_rate}
\end{figure}

\REVIEW{
\subsubsection{Spatio-temporal convergence rate of the low Mach enthalpy method} \label{sec_stefan_grid_convergence_study}

This section investigates the spatio-temporal convergence rate of the proposed low Mach enthalpy method. A grid convergence study is conducted for the three Stefan problems of Secs.~\ref{sec_stefan_novolchange}-~\ref{sec_stefan_shrinkage} using four grid sizes: $N_x  \times N_y = \{320 \times 16, 640 \times 32,1280 \times 64, 2560 \times 128\}$. A uniform time step size of $\Delta t=4\times10^{-4}$ s is employed for the coarse grid $N_x  \times N_y =320 \times 16$ and for each successive grid, the time step size is halved. This ensures the CFL number stays below 0.5 for all four grids. The temperature interval between liquidus and solidus is taken to be $\Delta T = 10$ K, as determined by the results of the previous section. The $\mathcal{L}^2$ error for a quantity $\psi$ is defined to be the root mean squared error (RMSE) of the vector $||\mathcal{E}_\psi ||_{\rm RMSE} =||\psi_\text{reference}-\psi_\text{numerical}||_2/\sqrt{\mathcal{N}}$. Here, $\mathcal{N}$ denotes the size of the vector $\mathcal{E}_\psi$. Two different reference solutions are considered here: (i) analytical and (ii) numerical solutions obtained using the finest grid resolution ($2560 \times 128$). Errors based on analytical solutions indicate the convergence of the present diffuse interface approach to its sharp interface counterpart. This is when the thickness of the mushy zone decreases. In contrast, errors based on the finest grid solutions indicate the spatio-temporal convergence rate of the diffuse interface model itself. We present errors as a function of mesh resolution for the interface position $x^*=s(t)$ for the entire simulation period ($0 \le  t \le 10$ s) and for temperature $T(x,t)$ in the entire domain ($0 \le x \le l$) at $t=5$ s. The uniform velocity in the liquid domain is a scalar multiple of the interface position (see Eq.~\eqref{eq_ul_u*}). Therefore, error in the interface position is also a measure of error in the flow field. 

Fig.~\ref{fig_ex0_convergence_rate} illustrates the spatio-temporal convergence rate of the numerical solution to the Stefan problem for the matched density case ($\rhol=\rhos$). Convergence rates based on the analytical solution are shown in Fig.~\ref{fig_ex0_convergence_rate} (A). As can be observed in the figure the present method exhibits close to first order with respect to the temperature solution. In the case of interface position, the present method exhibits a convergence rate between first and second order. We note that the interface location in diffuse interface enthalpy methods is implicitly defined as the 0.5 contour of the liquid fraction variable $\varphi$. Extracting the interface location $x^*$ itself introduces an unavoidable interpolation error that also contributes to the non-uniform convergence rate of $\mathcal{E}_{x^*}$. In this work we rely on VisIt software's \cite{childs2012visit} excellent post-processing capabilities to extract the interface location from the (distributed memory parallel) $\varphi$ data. Fig.~\ref{fig_ex0_convergence_rate} (B) illustrates the convergence rates based on the finest grid numerical solutions of $x^*$ and $T$. Here, the convergence rate between first and second order is observed for both quantities.

Figs.~\ref{fig_ex1_expansion_convergence_rate} (A) and (B) present the convergence rates for the Stefan problem exhibiting material expansion upon solidification ($R_\rho < 1$) using analytical and finest grid numerical solutions, respectively. The temperature convergence rate trend is the same as in the constant density case: close to first order convergence rate with respect to the analytical solution and between first and second order with respect to the finest grid numerical solution. Errors for interface position exhibit a non-monotonic convergence rate, however error magnitudes are low (on the order of $10^{-3}$). For larger temperature intervals ($\Delta T > 10$ K) the convergence rate of the interface position error is slightly better, but the error magnitude is higher (data not presented for brevity).

Finally, the method's accuracy is tested for the Stefan problem exhibiting material shrinkage upon solidification ($R_\rho > 1$) and the results are presented in Fig.~\ref{fig_ex1_shrinkage_convergence_rate}. Similar to the previous two cases, we observe close to first order convergence rate of temperature errors with respect to the analytical solution. We also observe close to second order convergence rate with respect to finest grid numerical solution. Interface position errors, though small saturate at fine grids.
}

\subsubsection{Solution to the transcendental equation}  \label{sec_sol_trans_eqn}
\begin{figure}
\centering
\includegraphics[width=0.9\linewidth]{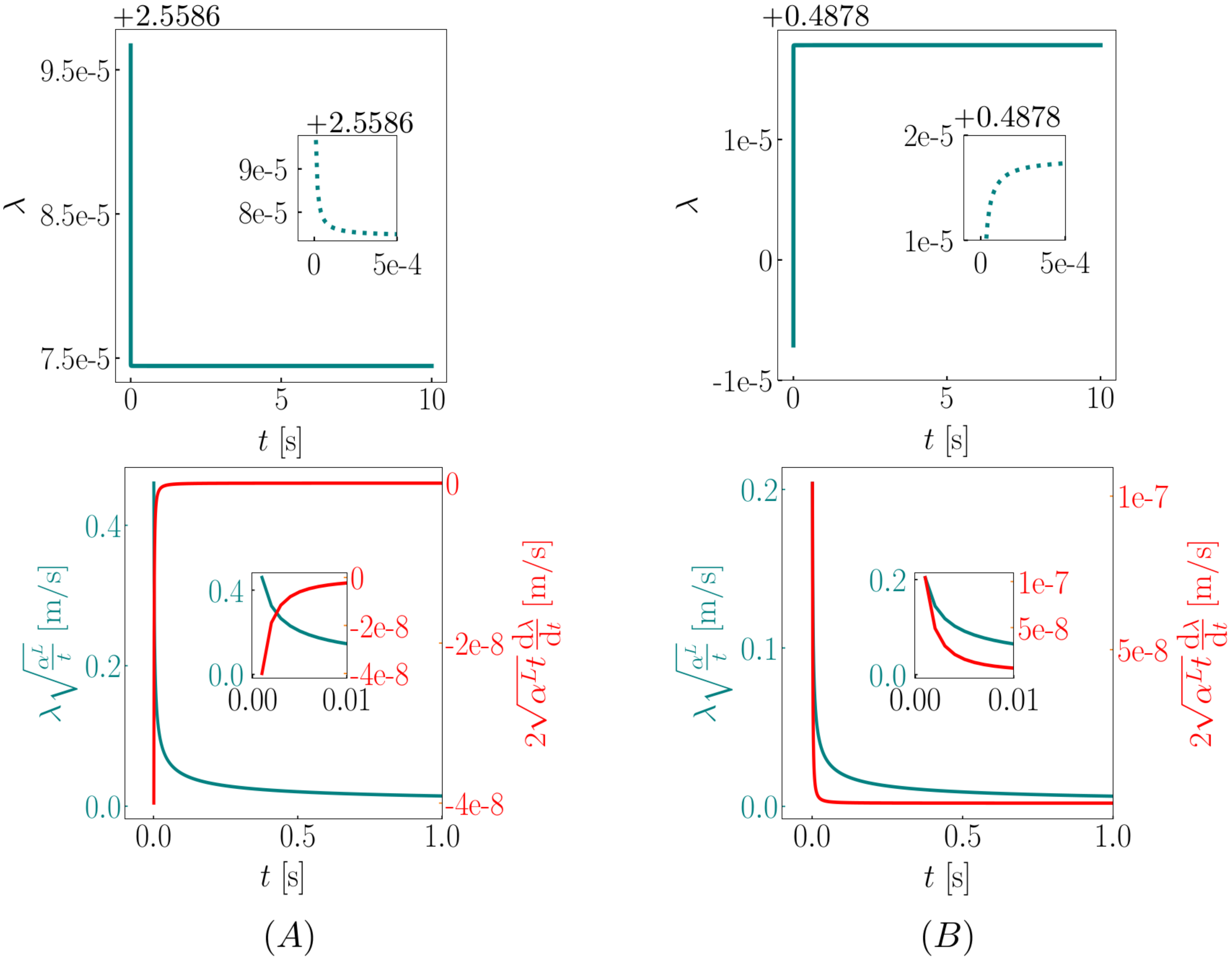}
\caption{\REVIEW{Variation of $\lambda$ and interface speed's two components $\lambda \sqrt{\frac{\alpha^{\rm L} }{t}}$ and $2 \sqrt{\alpha^{\rm L} t} \frac{\textrm{d}\lambda}{\textrm{d}t}$ (see Eq.~\ref{eq_dsdt})  as a function of time for the Stefan problem with (A) volume expansion and (B) volume shrinkage. $\lambda(t)$ is obtained by solving the transcendental equation (Eq.~\ref{eq_transcendental}) using MATLAB's \texttt{fzero} function. The transcendental equation is solved at $t = \Delta t$ and onwards. 
 $\frac{\textrm{d}\lambda}{\textrm{d}t}$ is computed from $\lambda$ in a post-processing step. }}
\label{fig_lambda}
\end{figure}

In Fig.~\ref{fig_lambda}, we plot $\lambda$ versus $t$ that is obtained by solving the transcendental Eq.~\eqref{eq_transcendental} for the expansion and shrinkage cases considered in this section. It is evident from the plot that $\lambda$ varies during the early stages of solidification (when kinetic energy dominates in the Stefan condition) before reaching a steady state.  \REVIEW{The relative magnitude of two terms comprising the interface velocity $\d s/ \d t$ (see Eq.~\ref{eq_dsdt}) are also compared for the expansion and shrinkage case in Fig.~\ref{fig_lambda}. We can observe that the second term involving $\d \lambda/ \d t$ is much smaller (at least six orders of magnitude) than the first term $\lambda/\sqrt{t}$, so it is justified to solve the simpler Eq.~\eqref{eq_transcendental} rather than the original, more complex Eq.~\eqref{eq_lambda_differential} for $\lambda(t)$.} For the thermophysical properties considered in this work, $\lambda$ variation is quite small and can arguably be ignored. Nevertheless, it is possible to include the kinetic energy jump term in the analytical solution to the two-phase Stefan problem. 

\subsubsection{Pressure jump across the interface for the Stefan problem exhibiting volume change}\label{sec_stefan_pr_jump}

\begin{figure}
\centering
\includegraphics[width=1.0\linewidth]{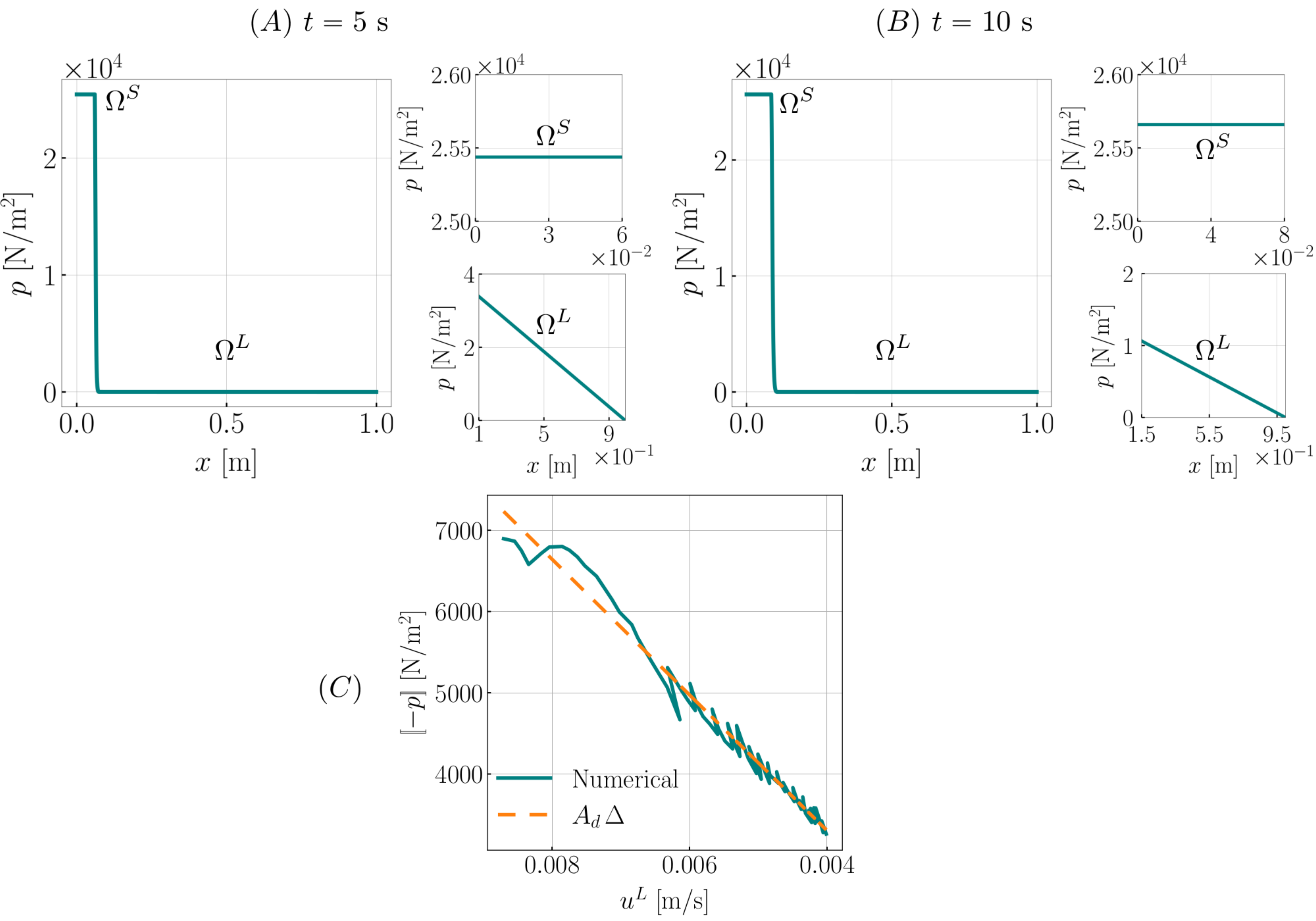}
\caption{Stefan problem with volume expansion ($R_\rho = 0.185$): Pressure distribution along the length of the channel and zoomed-in views for the liquid and solid domains at (A) $t=5$ s and (B) $t=10$ s. (C) Plot of the numerical pressure jump ($\llbracket -p \rrbracket = \ps - \pl$) across the interface as a function of liquid velocity $u^L$. The dashed line has a slope of value $A_d \Delta$, in which the Carman-Kozeny drag coefficient $A_d$ is computed using a solid fraction value of $\varphi_{\rm S} = 0.833$. We remark that for the purposes of this plot only the temperature interval is taken to be $\Delta T=60$ K. This is done to obtain a relatively smoother $\llbracket -p \rrbracket$ versus $\ul$ curve. Using $\Delta T = 10$ K, pressure jump across the mushy region exhibited larger oscillations as a function of fluid velocity. The grid size and time step size used for this case are $N_x \times N_y = 1280 \times 64$ and $\Delta t=10^{-4}$ s, respectively. }
\label{fig_stefan_pr_jump}
\end{figure}

Based on the analytical solution of the one-dimensional Stefan problem considering fluid flow, pressure varies linearly in the liquid phase and remains uniform in the solid phase. The numerical solution also exhibits this behavior of pressure variation.     Figs.~\ref{fig_stefan_pr_jump}(A) and \ref{fig_stefan_pr_jump}(B) show pressure in the entire domain at $t = 5$  and 10 s, respectively for the expansion case ($R_\rho = 0.185$). Zoomed-in plots are required to discern variation in liquid pressure since solid pressure is much greater. Although the numerical and analytical models predict the same trend in pressure variation, the numerical values differ substantially; numerical pressure values are much larger than the analytical ones (data not shown for brevity). This is due to the diffuse-interface formulation of the EM. Specifically, a Carman-Kozeny drag model is used in the EM to enforce no flow in the solid phase. In a diffuse formulation, velocity changes continuously from zero to a finite value within the mushy region. The pressure jump across the mushy region helps the fluid to ``squeeze" through. This is similar to the Darcy-Brinkman model of flow through porous regions \[ \u \propto -\grad p. \] The numerical pressure jump $\llbracket p \rrbracket = \pl - \ps$ ($\sim \grad p$) across the mushy region is plotted as a function of liquid velocity $\ul$. When the flow has subsided and the Darcy-Brinkman model becomes applicable, the curve is shown for $t > 2$ s.  There is a linear relationship between $\llbracket p \rrbracket$ and $\ul$, confirming our hypothesis that the numerical pressure jump occurs to push fluid through the mushy region. Additionally, the diffuse-domain momentum equation provides a magnitude scale of $\llbracket p \rrbracket$ 
\begin{align}
-\D{p}{x} & \sim A_d \, \ul \nonumber \\
\hookrightarrow \ps - \pl  =  \llbracket -p \rrbracket &\sim A_d  \, \Delta \, \ul,  \label{eq_slope_dp}  
\end{align} 
in which $\Delta$ is the cell size and $A_d(\varphi_{\rm S})$ is the drag coefficient. Here, we have ignored the convective scale $(\rho \D{\ul}{t})$ in the mushy zone as it is several orders lower than the pressure gradient and drag force terms (data not shown). Eq.~\eqref{eq_slope_dp} suggests that the slope of $\llbracket -p \rrbracket$ versus $\ul$ curve is  $A_d \Delta$. This is confirmed from Fig.~\ref{fig_stefan_pr_jump} where a line of slope $A_d \Delta$ evaluated at a solid fraction $ \varphi_{\rm{S}} = 0.833$ \REVIEW{(this value of $\varphi_\text{S}$ is obtained by equating the slope of the best-fit linear curve to $A_d \Delta$, and then solving for $\varphi_\text{S}$)} captures the $\llbracket -p \rrbracket$ versus $\ul$ trend reasonably well. 

\subsection{Metal melting} \label{sec_metal_melting}
\begin{figure}
\centering
\includegraphics[width=0.8\linewidth]{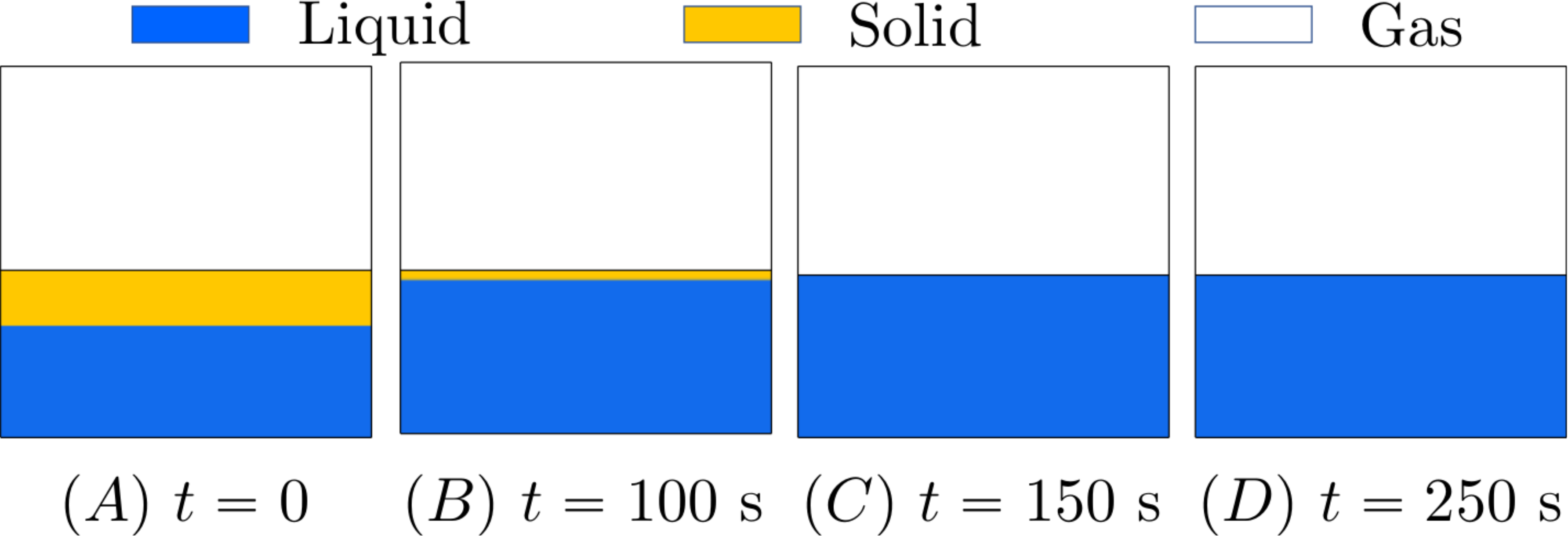}
\caption{Time evolution of the solid, liquid, and gas domains during metal melting.}
\label{fig_mass_conservation}
\end{figure}
As our next example\footnote{\REVIEW{This benchmark test is provided in IBAMR GitHub within the directory \texttt{examples/phase\_change/ex3}.}}, we simulate melting of aluminum metal with a free surface to highlight two salient features of the new low Mach enthalpy method: (\textbf{1}) the ability to capture volume change effects of the PCM in the presence of gas phase; and (\textbf{2}) the ability to handle phase appearance or disappearance from the domain.  This problem is inspired by Huang et al.~\cite{huang2022consistent} who solved a similar problem using a phase-field method (PFM), but using hypothetical\footnote{As mentioned earlier it is difficult to estimate/measure various material properties that is required in a PFM.} thermophysical properties of the PCM.  The computational domain is considered to be a unit square $\Omega \in [0, 1]^2$ that is discretized by $N \times N = 256 \times 256$ grid cells. At $t = 0$ the heavier liquid phase of density $\rhol = 2700$ kg/m$^3$ occupies the region below $y = 0.3$ m, and the lighter solid phase of density $\rhos = 2475$ kg/m$^3$ rests above the liquid phase and fills the domain until $y = 0.45$ m; see Fig.~\ref{fig_mass_conservation} (A). The gas occupies the remaining domain ($0.45 < y <= 1$). Both liquid and solid phases are assumed to have the same viscosity $\mul=\mus=1.4\times10^{-3}$ kg/m$\cdot$s. Viscosity in the solid phase is fictitious and does not affect numerical results. The rest of the thermophysical properties for the liquid and solid phases are taken from  Table~\ref{tab_aluminum_properties}.  The thermophysical properties of the  gas are based on air and are taken to be  $\rhog=0.4$ kg/m$^3$, $\kg=6.1\times10^{-2}$ W/m$\cdot$K, $\cpg=1100$ J/kg$\cdot$K  and $\mug=4\times10^{-5}$ kg/m$\cdot$s. In the $x$-direction, periodic boundary conditions are applied. On the top boundary ($y = 0$),  we use zero-pressure/outflow and no heat flux boundary conditions, while on the bottom wall ($y = 0$), we use zero-velocity and fixed temperature boundary conditions ($T = 6T_m$). $T_m = 933.6$ K is the melting temperature of aluminum. For the liquid region, the initial temperature is $5T_m$, whereas for the solid and gas regions, it is $0.9T_m$. The simulation is performed till $t=250$ s with a uniform time step size of $\Delta t = 10^{-3}$ s.

The solid melts when heat is transferred from the bottom wall to the liquid. As the melting process continues, the solid phase disappears after $t = 150$ s.  To ensure that no spurious phase changes or interfacial dynamics exist, the simulation is continued until $t = 250$ s. As shown in panels (C) and (D) of Fig.~\ref{fig_mass_conservation}, this is indeed the case. Since liquid density is larger than solid density, the gas-liquid interface position ($y = 0.4375$ m) at $t = 150$ s is lower than the initial gas-solid interface position ($y = 0.45$ m). It provides a simple ``sanity check" for a CFD method that seeks to capture the volume change effect of the PCM; the final position of the gas-liquid interface indicates the success of the method. Additionally, the initial and final gas-metal interface locations can be used to quantify the percentage change in metal mass. It should ideally be zero. For the present simulation (using $256 \times$ 256 grid), it is approximately $\mathcal{E} \approx 0.027 \%$. Even though the percentage mass change is quite small, it is not near machine precision. This is attributed to two factors: (\textbf{1}) the non-conservative nature of the level set method, which is used to track the gas-PCM interface in our formulation; and (\textbf{2}) the errors incurred in computing the right-hand side of the low Mach Eq.~\eqref{eq_divu} numerically. We show in Appendix Sec.~\ref{sec_mass_conservation_convergence} that $\mathcal{E}$ decreases with increasing grid resolution. It also decreases with decreasing time step size $\Delta t$ (data not presented). We expect mass change errors would be lower if the level set method were replaced by a more conservative interface capturing technique like the volume of fluid technique. However, this needs to be verified.

\subsection{Metal solidification} \label{sec_metal_solidification}
We consider the opposite scenario of the previous section in our final example: molten aluminum solidifying in a cast\footnote{\REVIEW{This benchmark test is provided in IBAMR GitHub within the directory \texttt{examples/phase\_change/ex4}.}}. A common casting defect is solidification shrinkage. Metals that are denser in their solid form than in their liquid form (which is almost always the case) suffer from this defect. As liquid metal solidifies, its volume contracts due to the density contrast between phases. Macroscopically, the free surface of the metal is recessed down into the solidified metal, giving it the appearance of a pipe. The goal of this example is to demonstrate that pipe defects are captured only when velocity is taken to be non-div-free, and the prior inconsistent enthalpy methods \cite{yan2018fully,lin2020conservative,panwisawas2017mesoscale} would not be able to capture this feature of phase change process. In those works, a divergence-free (div-free) velocity condition is used, which means that no additional flow is generated when the solid phase changes to liquid or vice versa. In the momentum and energy equations, the two phases (liquid and solid) are allowed to have different densities, which is at odds with the div-free velocity assumption. 

\begin{figure}
\centering
\includegraphics[width=0.8\linewidth]{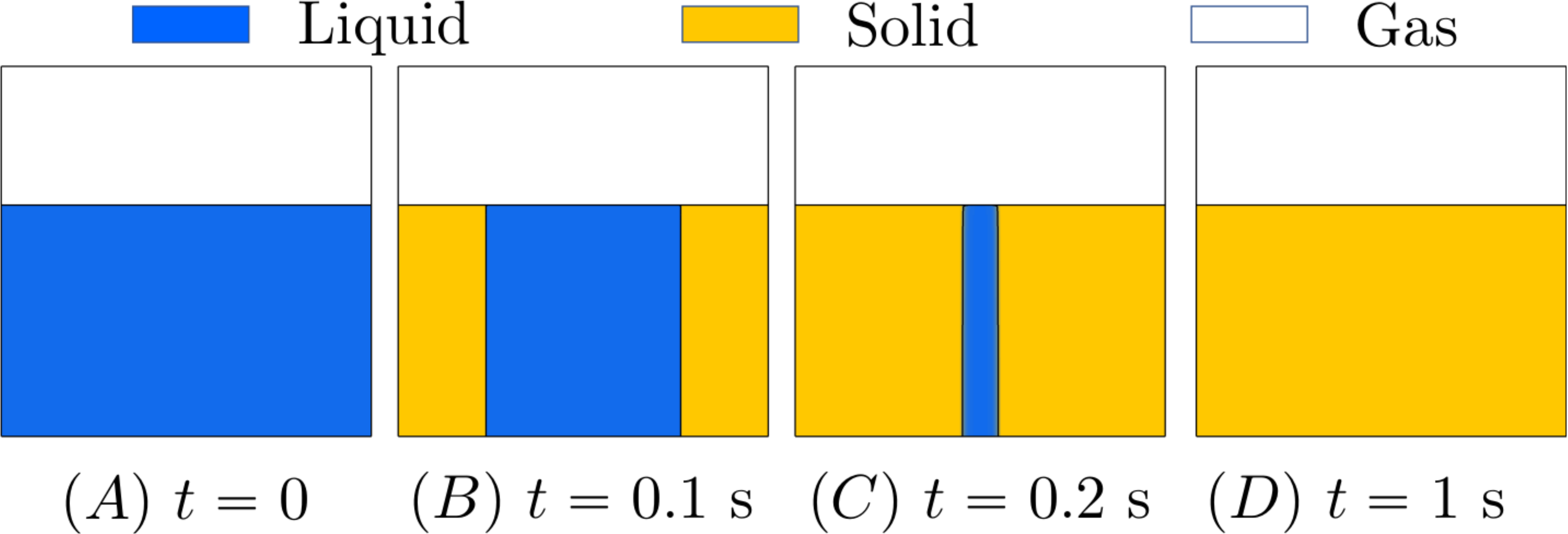}
\caption{Time evolution of the solid, liquid, and gas domains during metal solidification considering $\div \u = 0$.}
\label{fig_pipe_shrinkage_wo_div_u}
\end{figure}

Solidification of aluminum occurs within a square computational domain of extents $\Omega \in [0, 8\times10^{-3}]^2$, which is discretized into $256\times256$ uniform cells. Initially, liquid metal is filled below the region $y=5\times10^{-3}$ m, and the rest of the domain is occupied by gas (air); see Fig.~\ref{fig_pipe_shrinkage_wo_div_u}(A). In contrast to the previous example, there is no solid phase at the beginning of the simulation. The densities of liquid and solid are assumed to be  $\rhol = 2475$ kg/m$^3$ and $\rhos = 2700$ kg/m$^3$, respectively. The other thermophysical properties for liquid, solid, and gas are taken from the previous section.  The surface tension coefficient between liquid aluminum and gas is taken to be $\sigma = 0.87$ N/m.  On the top surface, zero-pressure/outflow boundary conditions are applied, while zero-velocity boundary conditions are applied elsewhere. On all boundaries, the temperature is fixed at $T=0.5T_m$ ($T_m = 933.6$ K is the solidification temperature of aluminum), except at the bottom wall, where a zero-flux (homogenous Neumann) condition is imposed. The initial temperature in the liquid domain is set to $2T_m$, whereas in the gas domain it is $0.5T_m$. As a result of the imposed boundary conditions, solidification begins on the right and left sides of the domain. Solidification is not affected by the top boundary since air has a low thermal conductivity. The simulation runs until $t=1$ s with a uniform time step size of $\Delta t = 10^{-5}$ s. 

\begin{figure}
\centering
\includegraphics[width=0.8\linewidth]{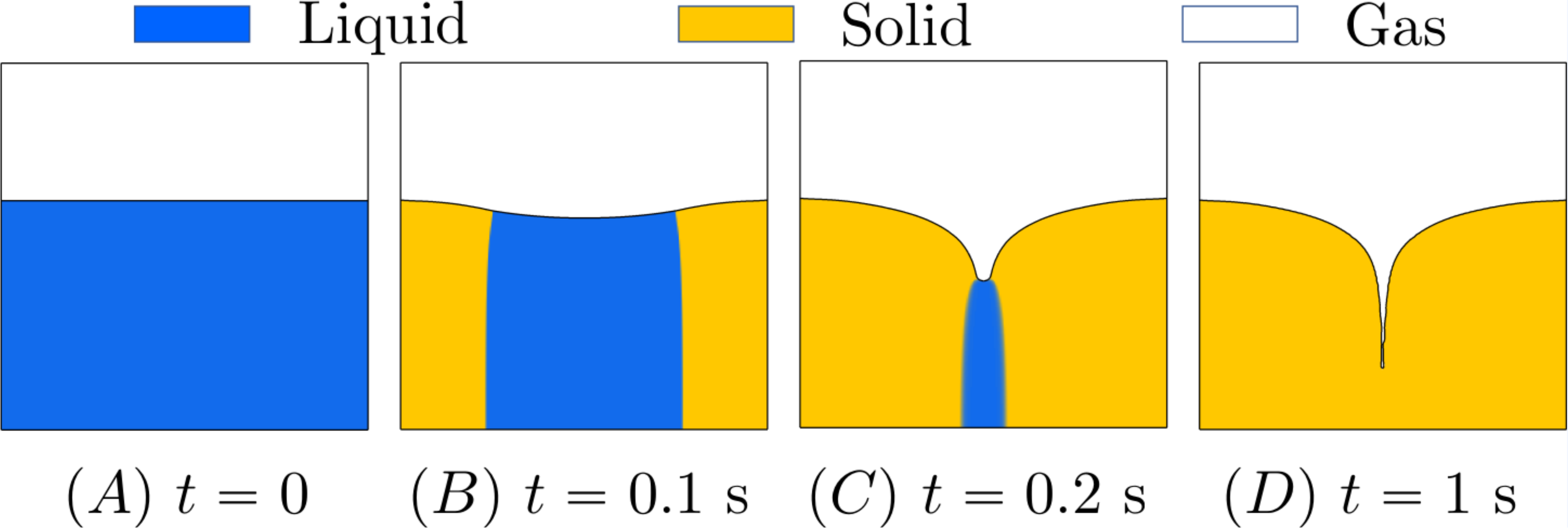}
\caption{Time evolution of the solid, liquid, and gas domains during metal solidification (shrinkage) considering $\div \u \ne 0$.}
\label{fig_pipe_shrinkage}
\end{figure}

First, we present the simulation results assuming that velocity in the domain is div-free. Results are shown in  Fig.~\ref{fig_pipe_shrinkage_wo_div_u}. In this case, the liquid metal solidifies completely without deforming the free surface or changing its volume. This solution is unphysical because PCM volume must change to accommodate a change in density during phase change. It is important to note that variable density and gravitational body forces are taken into account in the momentum equation \eqref{eq_momentum}. However, this is not enough to cause caving of the free surface (gas-liquid).    

The results differ significantly when velocity is not treated as div-free. Fig.~\ref{fig_pipe_shrinkage} shows the dynamics of solidification.  Around $t=0.25$ s, the liquid metal solidifies completely, and no further phase changes occur. There are no spurious flows or interfacial motions at the gas-solid interface beyond $t = 0.25$ s. It is clear that the gas-liquid surface has caved and a pipe defect has formed. Material volume shrinks as density increases, so this is a physically correct result. In addition, we quantify the simulation results by computing the percentage mass change of the metal as a function of grid resolution. This is presented in the Appendix Sec.~\ref{sec_mass_conservation_convergence}. 

\begin{figure}
\centering
\includegraphics[width=0.8\linewidth]{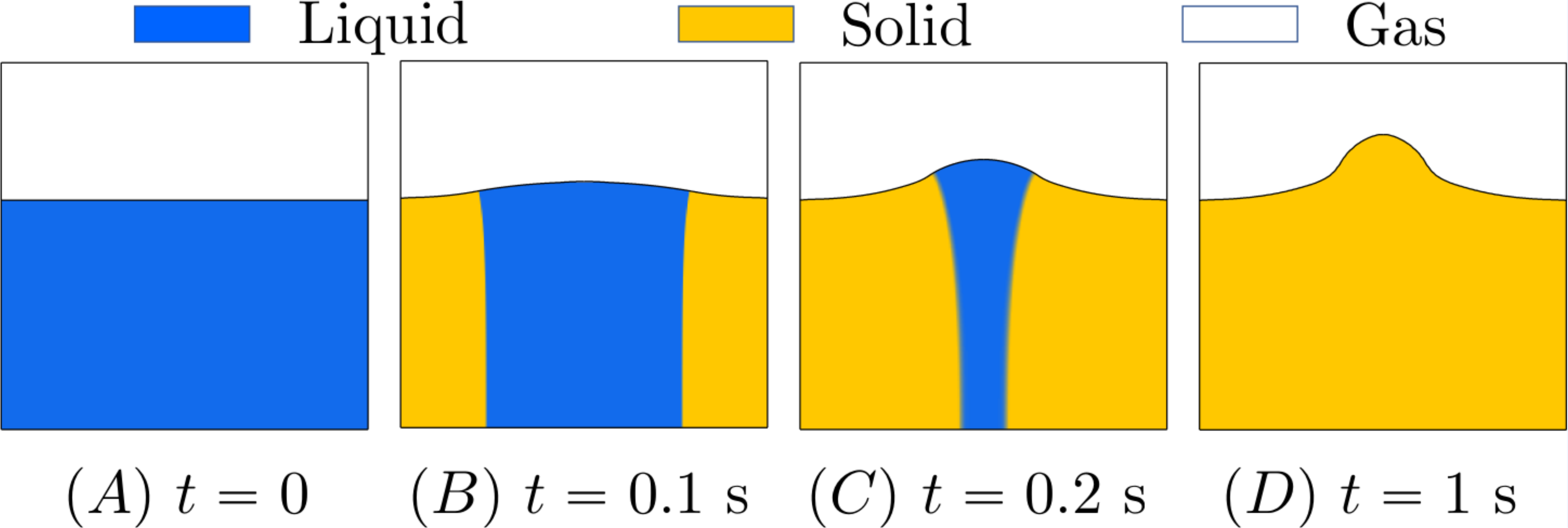}
\caption{\NEW{Time evolution of the solid, liquid, and gas domains during metal solidification (expansion) considering $\div \u \ne 0$.}}
\label{fig_pipe_expansion}
\end{figure}

\NEW{Next, consider a hypothetical case where aluminum liquid and solid phase densities are reversed to $\rhol=2700$ and $\rhos=2475$ kg/m$^3$, respectively. As a result, metal expands during solidification. Fig.~\ref{fig_pipe_expansion} shows the dynamics of metal expansion during solidification. It takes $t=0.25$ s for solidification to complete in this case as well, but we observe a protrusion defect instead of a pipe defect.} 

\section{Discussion} 
In this study we presented analytical and numerical methodologies to model phase change phenomena exhibiting density/volume changes. All materials of practical interest change density, and consequently, material volume to conserve mass when they melt, solidify, evaporate, or condense. Different phases of the same material also exhibit differences in other thermophysical properties, such as specific heat and conductivity.  By retaining all jump terms arising from the energy equation in the Stefan condition, we derived an analytical solution to the two-phase Stefan problem with variable thermophysical properties. For validation purposes, CFD algorithms that aim to model the phase change of materials can benefit greatly from the two-phase Stefan problem solutions, but these have gone largely unnoticed in the literature.  We also presented a novel low Mach version of the enthalpy method that takes into account the density change of PCMs during melting and solidification. Furthermore, the solid-liquid PCM was coupled to a gas phase within the low Mach framework. Evaporation and condensation may also be modeled with the proposed low Mach enthalpy method, but further studies are needed to evaluate its accuracy. Another possibility is to model evaporation and condensation through the level-set or volume of fluid machinery and melting and solidification via the proposed low Mach enthalpy method. In this scenario $\DDD{H}{t}$ and $\DDD{\phi}{t}$ are not equal to zero in the low Mach Eq.~\eqref{eq_divu}.  By using such a framework, all four modes of phase change can be handled in a single simulation, which will enhance the existing modeling fidelity of engineering applications like metal additive manufacturing.  Besides presenting novel techniques for modeling phase change processes, this work opens up several new directions for future research.   

\section*{Acknowledgements}
R.T. and A.P.S.B~acknowledge support from NSF awards OAC 1931368 and CBET CAREER 2234387. Compute time on SDSU's high performance computing cluster Fermi is greatly acknowledged.


\appendix
\renewcommand\thesection{\Alph{section}}
\input{Appendix}

\newpage
\section*{Bibliography}
\begin{flushleft}
 \bibliography{stefan_bibliography.bib}
\end{flushleft}

\end{document}

%% file: Appendix.tex
\section{Similarity solution} \label{sec_similarity_soln}
Here, we derive a similarity solution of the heat equations governing temperature in the solid $\Omegas$ and liquid $\Omegal$ domains

\begin{align}
&\frac{\partial \Ts}{\partial t} =\alphas\frac{\partial^2 \Ts}{\partial x^2} \qquad \qquad \qquad  \qquad \quad \quad \in  \Omegas(t),  \label{eq_temp_solid3}  \\
&\frac{\partial \Tl}{\partial t} + \left( 1 - R_\rho \right) \frac{\d s}{\d t} \frac{\partial \Tl}{\partial x} =\alphal\frac{\partial^2 \Tl}{\partial x^2} \quad \; \;  \in \Omegal (t). \label{eq_temp_liquid3}
\end{align}
Interface position $s(t)$ can be expressed as $s(t) =  \lambda(t) 2\sqrt{\alphal t}$, where $\lambda(t)$ is an unknown function of time. Eqs.~(\ref{eq_temp_solid}) and (\ref{eq_temp_liquid}) can be written in terms of similarity variables $\etas$ and $\etal$, respectively:
\begin{align}
\Ts(x,t) &= \Ts(\etas) \qquad \text{with} \qquad \etas = \frac{x}{2\sqrt{\alphas t}}     \label{eq_Ts_sim},  \\
\Tl(x,t) &= \Tl(\etal) \qquad \text{with} \qquad \etal = \frac{x}{2\sqrt{\alphal t}} + b(t)   \label{eq_Tl_sim}.
\end{align}
$b(t)$ in $\etal$ is yet to be determined. The similarity transformation reduces partial differential equations (\ref{eq_temp_solid}) and (\ref{eq_temp_liquid}) in $x$ and $t$ to ordinary differential equations in $\etas$ and $\etal$, respectively. The steps involved in the similarity transformation of \eqref{eq_temp_liquid} are detailed in the remainder of this section as this equation is different from the standard heat~\eqref{eq_temp_solid}, which has been treated in several textbooks. The main steps involve rewriting the derivatives of $\Tl$ and $s(t)$

\begin{align}
\D{\Tl}{x} = \dd{\Tl}{\etal} \D{\etal}{x} = \dd{\Tl}{\etal}  \frac{1}{2\sqrt{\alphal t}}  \\
\DD{\Tl}{x} =  \frac{\rm d^2 \Tl}{\rm d {\etal}^2} \D{\etal}{x}  \frac{1}{2\sqrt{\alphal t}} =  \frac{\rm d \Tl}{\rm d^2 {\etal}^2} \frac{1}{4 \alphal t} \\
\D{\Tl}{t} = \dd{\Tl}{\etal} \D{\etal}{t} = \dd{\Tl}{\etal}  \left(\frac{-x}{4t\sqrt{\alphal t}} + \dd{b}{t}\right) \\
\dd{s}{t} = \lambda \sqrt{\frac{\alphal}{t}} +  2\sqrt{\alphal t} \; \dd{\lambda}{t}
\end{align}
and substituting them in \eqref{eq_temp_liquid}. This yields
\begin{align}
  \frac{\rm d^2 \Tl}{\rm d {\etal}^2} + 2\left(\etal - b - \lambda (1 - R_\rho)  -2 t \dd{b}{t} -2 t (1 - R_\rho) \dd{\lambda}{t} \right) \dd{\Tl}{\etal} = 0. \label{eq_bt}
\end{align}
Here, $ \frac{\rm d^2 \Tl}{\rm d {\etal}^2} = \frac{\rm d} {\rm d \etal} \left(\dd{\Tl}{\etal}\right)$. Choosing $b(t) = -\lambda(t)(1 - R_\rho)$ in \eqref{eq_bt} simplifies the similarity transformation of  \eqref{eq_temp_liquid} to
\begin{align}
  \frac{\rm d^2 \Tl}{\rm d {\etal}^2} + 2\etal  \dd{\Tl}{\etal} = 0.  \label{eq_liquid_sim}
\end{align}
Following similar (but less involved) steps, the similarity transformation of \eqref{eq_temp_solid} reads as
\begin{align}
  \frac{\rm d^2 \Ts}{\rm d {\etas}^2} + 2\etas  \dd{\Ts}{\etas} = 0.  \label{eq_solid_sim}
\end{align}
Eqs.~(\ref{eq_liquid_sim}) and (\ref{eq_solid_sim}) can be integrated analytically and the closed-form expressions of $\Ts(x,t) = \Ts(\etas)$ and $\Tl(x,t) = \Tl(\etal)$ (satisfying boundary and interface conditions) are provided in the main text.

\REVIEW{\section{Spatial discretization} \label{sec_spatial_discretization}
In this work, a staggered Cartesian grid is used to discretize the continuous equations of motion. The computational domain $\Omega$ is discretized into uniform $N_x\times N_y$ cells with grid spacing of $\Delta x$ and $\Delta y$ in each direction as shown in Fig.~\ref{fig_discretized_staggered_grid}. The bottom left corner of the computational domain $\Omega$ is assumed to align with the origin $(0,0)$.  The position of each grid cell center is then given by $\x_{i,j} = \left((i + \half)\dx,(j + \half)\dy\right)$, where $i = 0, \ldots, \Nx - 1$ and $j = 0, \ldots, \Ny - 1$. The face center in the $x-$direction which is half grid space away from the cell center $\x_{i,j}$ in negative $x$ direction is given by $\x_{i-\half,j} = \left(i\dx,(j + \half)\dy\right)$, where $i = 0, \ldots, \Nx$ and $j = 0, \ldots, \Ny - 1$. Other face center physical locations are analogous. The scalar variables such as pressure $p$, specific enthalpy $h$, and temperature $T$ are stored at the cell centers. The x-component of the velocity ($u$) is stored at the x-direction cell faces while the y-component of the velocities is stored at the y-direction cell faces, as shown in Fig.~\ref{fig_single_cell}. All the material properties including density are stored at the cell centers along with the interface tracking variables $\varphi$ and $\phi$. Second-order interpolation is used when the cell-centered quantities are sought at the face. The momentum forcing terms such as surface tension, gravity, and Carman-Kozeny drag force are stored at the cell faces as the velocity components. 
\begin{figure}
  \centering
  \subfigure[2D staggered Cartesian grid]{
    \includegraphics[scale = 0.45]{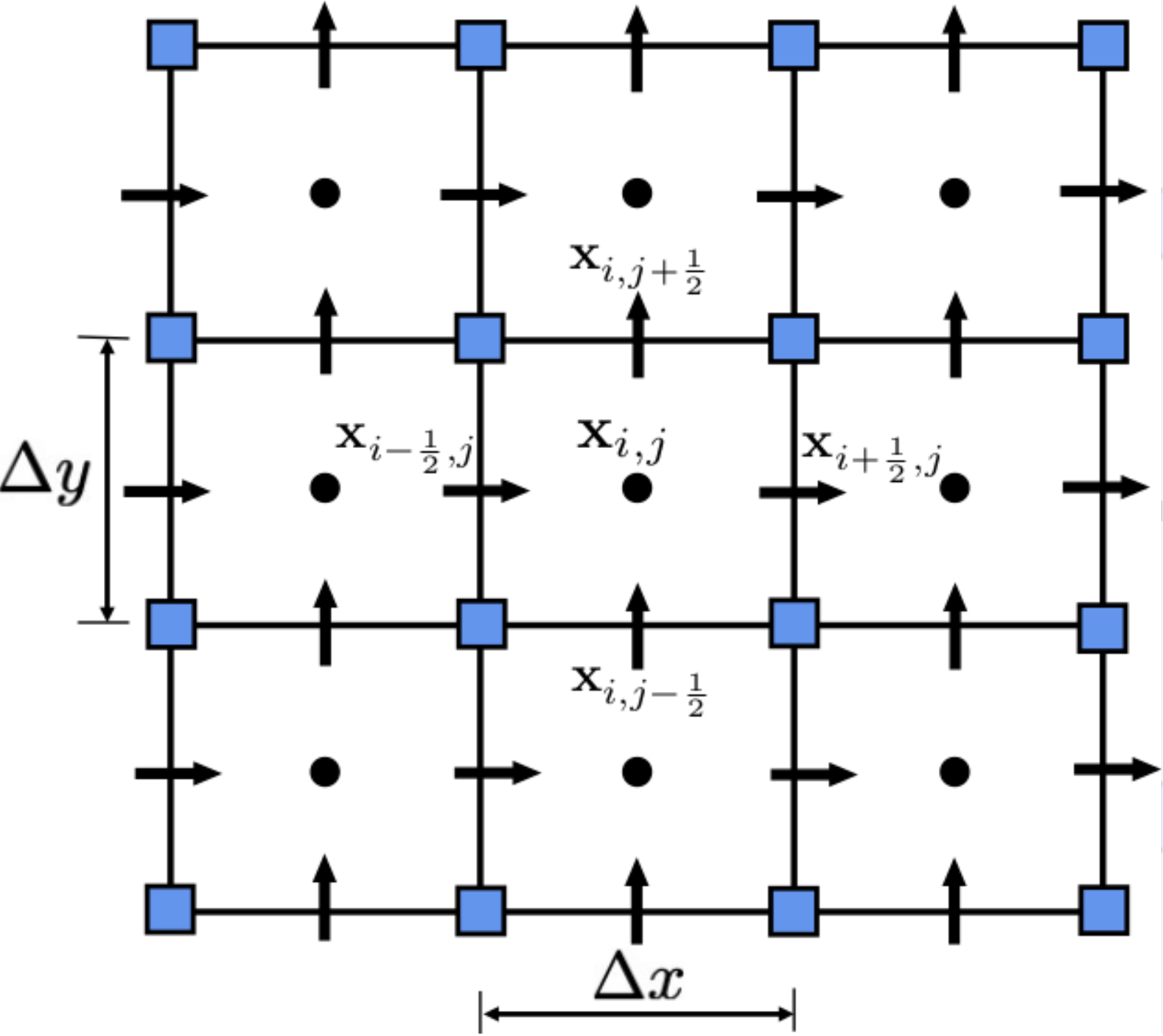} 
    \label{fig_discretized_staggered_grid}
  }
   \subfigure[Single grid cell]{
    \includegraphics[scale = 0.35]{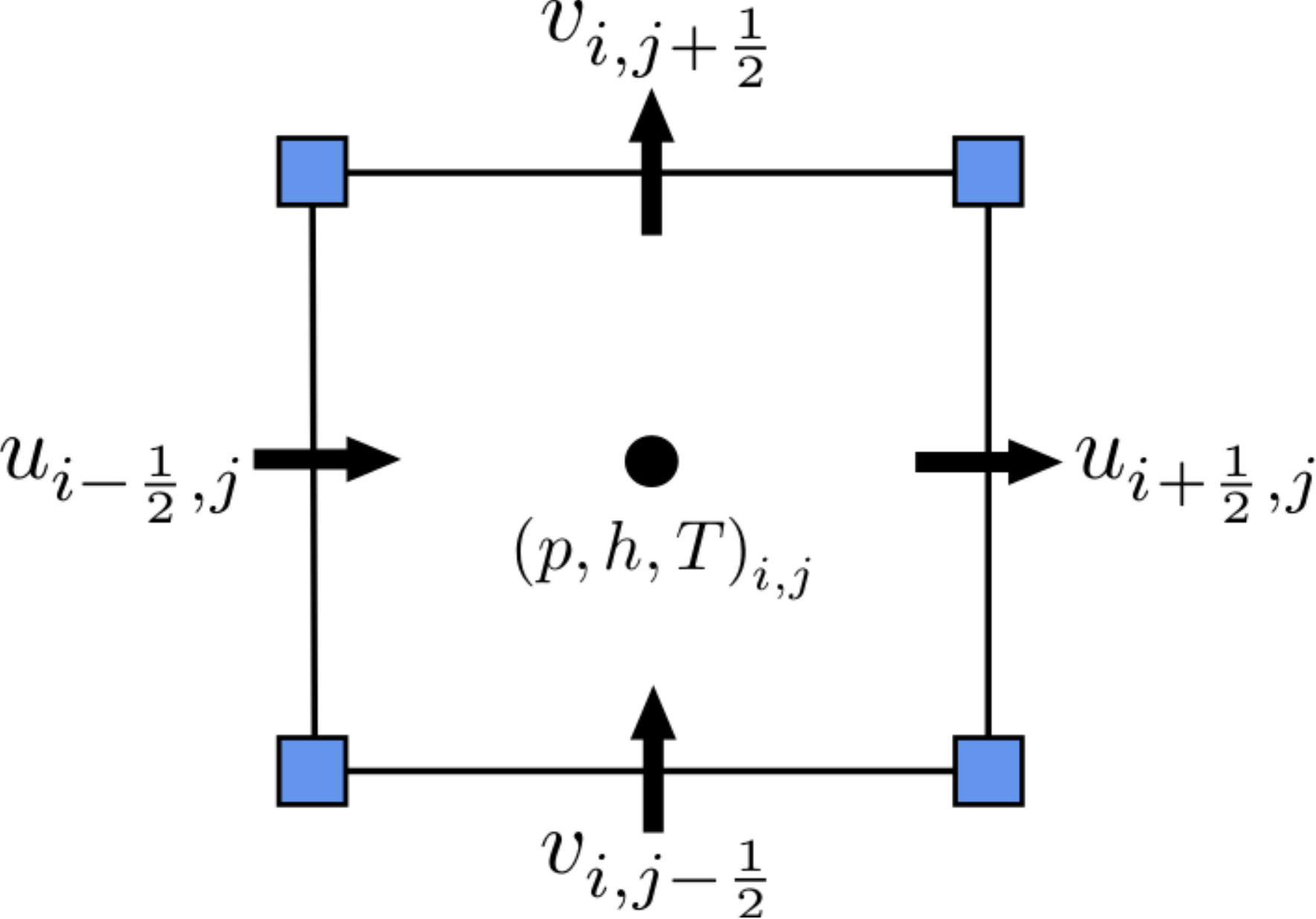}
    \label{fig_single_cell}
  }
  \caption{\REVIEW{Schematic representation of a 2D staggered Cartesian grid. \subref{fig_discretized_staggered_grid} shows the coordinate system for the staggered grid. \subref{fig_single_cell} shows a single grid cell with velocity components $u$ and $v$ approximated at the cell faces (${\bf{\rightarrow}}$) and scalar variables pressure $p$, specific enthalpy $h$ and temperature $T$ approximated at the cell center ($\bullet$).}}
\label{fig_cfd_grid}
\end{figure}
We use standard second-order finite differences to approximate the 
spatial differential operators. The following are the spatial discretizations of the key continuous operators: 
\begin{itemize}
\item The gradient of cell-centered quantities (i.e., $p$) is approximated at cell faces by
\begin{align}
\label{eq_grad_fd}
& \G p = (G^x p, G^y p), \\
&(G^x p)_{i-\half,j} = \frac{p_{i,j} - p_{i-1,j}}{\dx}, \\
&(G^y v)_{i,j - \half} =\frac{p_{i,j} - p_{i,j-1}}{\dy}. 
\end{align}
\item The continuous strain rate tensor form of the viscous term is
\begin{equation}
\label{eq_visc_cont}
\div \left[\mu \left(\grad \u + \grad \u^\intercal\right) \right] = 
\left[
\begin{array}{c}
 2 \D{}{x}\left(\mu \D{u}{x}\right) + \D{}{y}\left(\mu\D{u}{y}+\mu\D{v}{x}\right) \\
 2 \D{}{y}\left(\mu \D{v}{y}\right) + \D{}{x}\left(\mu\D{v}{x}+\mu\D{u}{y}\right) \\
\end{array}
\right],
\end{equation}
which couples the velocity components through spatially variable viscosity 
\begin{equation}
\label{eq_visc_discrete}
\Lmu \u= 
\left[
\begin{array}{c}
 (\Lmu \u)^x_{i-\half,j} \\
 (\Lmu \u)^y_{i,j-\half}  \\
\end{array}
\right].
\end{equation}
\item The viscous operator is discretized using standard second-order, centered finite differences
\begin{align}
 (\Lmu \u)^x_{i-\half,j} &= \frac{2}{\dx}\left[\mu_{i,j}\frac{u_{i+\half,j} - u_{i-\half,j}}{\dx} -
					        \mu_{i-1,j}\frac{u_{i-\half,j} - u_{i-\3half,j}}{\dx}\right] \nonumber \\ 
                    &+ \frac{1}{\dy}\left[\mu_{i-\half, j+\half}\frac{u_{i-\half,j+1} - u_{i-\half,j}}{\dy} - 
					         \mu_{i-\half, j-\half}\frac{u_{i-\half,j} - u_{i-\half,j-1}}{\dy}\right] \nonumber \\
	            &+ \frac{1}{\dy}\left[\mu_{i-\half, j+\half}\frac{v_{i,j+\half} - v_{i-1,j+\half}}{\dx} - 
					         \mu_{i-\half, j-\half}\frac{v_{i,j-\half} - v_{i-1,j-\half}}{\dx}\right] \label{eq_viscx_fd} \\				         
 (\Lmu \u)^y_{i,j-\half} &= \frac{2}{\dy}\left[\mu_{i,j}\frac{v_{i,j+\half} - v_{i,j-\half}}{\dy} -
					        \mu_{i,j-1}\frac{v_{i,j-\half} - v_{i,j-\3half}}{\dy}\right] \nonumber \\ 
                    &+ \frac{1}{\dx}\left[\mu_{i+\half, j-\half}\frac{v_{i+1,j-\half} - v_{i,j-\half}}{\dx} - 
					         \mu_{i-\half, j-\half}\frac{v_{i,j-\half} - v_{i-1,j-\half}}{\dx}\right] \nonumber \\
	            &+ \frac{1}{\dx}\left[\mu_{i+\half, j-\half}\frac{u_{i+\half,j} - u_{i+\half,j-1}}{\dy} - 
					         \mu_{i-\half, j-\half}\frac{u_{i-\half,j} - u_{i-\half,j-1}}{\dy}\right] \label{eq_viscy_fd},
\end{align}
in which viscosity is required on both cell centers and \textit{nodes} of the staggered grid (i.e., $ \mu_{i\pm\half, j\pm\half}$).
Node-centered quantities are obtained via interpolation by either arithmetically or harmonically averaging
the neighboring cell-centered quantities. In this work, we use harmonic averaging. 
\end{itemize}
All other equations such as energy, level-set and Heaviside advection, and low Mach are discretized at the cell centers as follows:
\begin{itemize}
\item The divergence of the velocity field
$\u = (u,v)$ is approximated at cell centers by
\begin{align}
\label{eq_div_fd}
& \vD \cdot \u = D^x u + D^y v, \\
&(D^x u)_{i,j} = \frac{u_{i+\half, j} - u_{i-\half, j}}{\dx}, \\
&(D^y v)_{i,j} = \frac{v_{i, j+\half} - v_{i, j-\half}}{\dy}. 
\end{align}
\item The diffusion term in the energy equation is approximated as
\begin{align}
\label{eq_energy_diff}
\vD \cdot \left(\kappa \grad{T}\right) = &\frac{1}{\dx}\left(\kappa_{i+\half,j}\frac{T_{i+1,j} - T_{i,j}}{\dx} - \kappa_{i-\half,j}\frac{T_{i,j} - T_{i-1,j}}{\dx}\right) \nonumber \\ 
					     &+\frac{1}{\dy} \left(\kappa_{i,j+\half}\frac{T_{i,j+1} - T_{i,j}}{\dy} - \kappa_{i,j-\half}\frac{T_{i,j} - T_{i,j-1}}{\dy}\right)
\end{align}
\end{itemize}
Harmonic average is employed to interpolate thermal conductivity $\kappa$ from cell centers to cell faces. Extending these discretizations to three-dimensional Cartesian grids is straightforward. For convective discretization, we use third-order accurate cubic upwind interpolation (CUI). CUI satisfies both the convection-boundedness criterion (CBC) and the total variation diminishing (TVD) property. The CUI scheme exhibits third-order spatial accuracy in monotonic regions (specifically where the gradient of the advected quantity remains monotone) and reduces to first-order spatial accuracy (due to upwinding) in non-monotonic regions. For brevity, we omit the spatial discretization of the advection equation using CUI, but it can be found in our previous work \cite{nangia2019robust}.
}
\section{Grid convergence study for mass conservation}\label{sec_mass_conservation_convergence}

\begin{figure}
\centering
\includegraphics[width=0.6\linewidth]{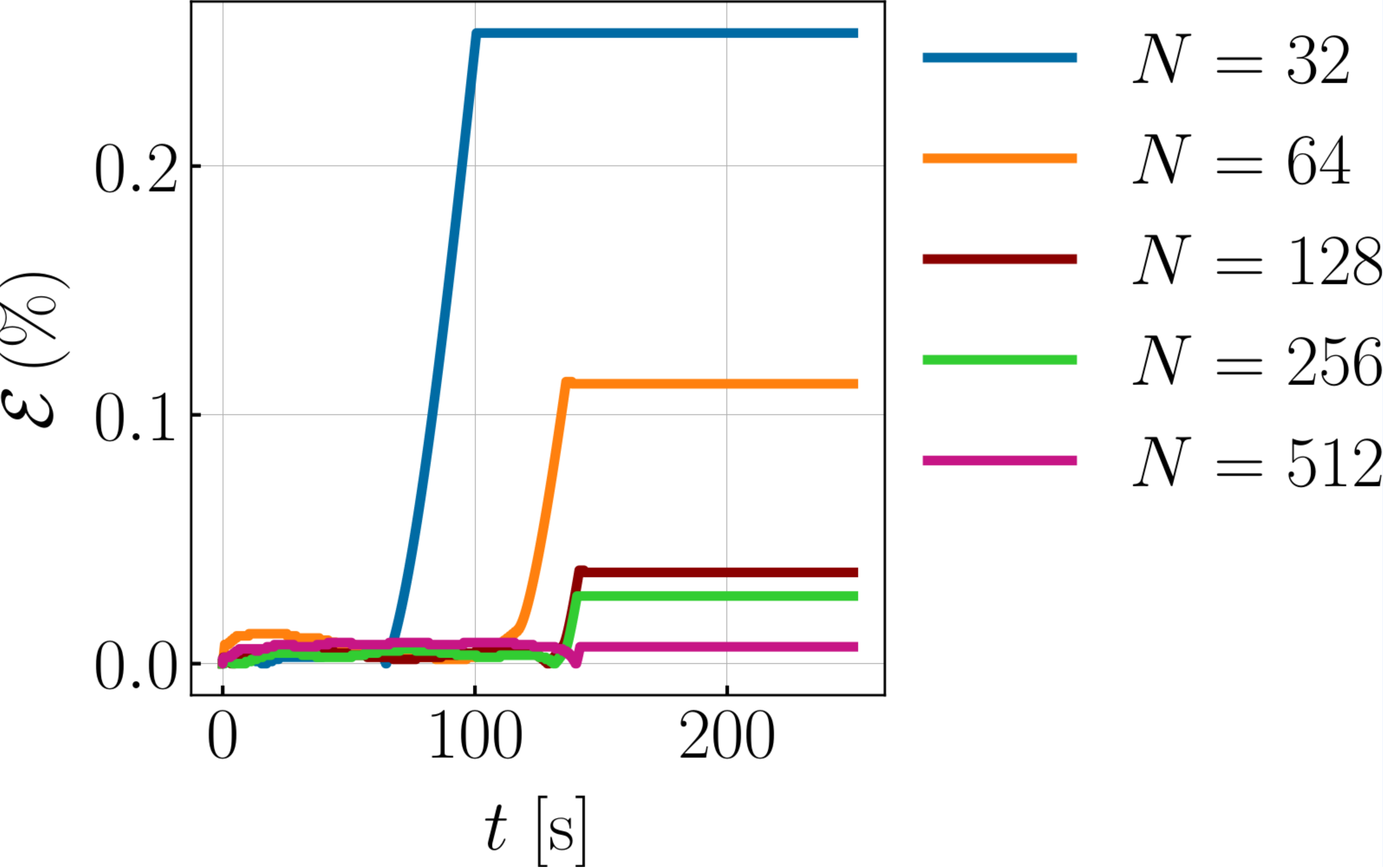}
\caption{Percentage change in PCM mass $\mathcal{E}$ as a function of time for the metal melting case at different grids.  Each grid uses a uniform time step size of $\Delta t=10^{-3}$ s and a temperature interval of $\Delta T=10$ K.}
\label{fig_mass_conservation_error}
\end{figure}

\begin{figure}
\centering
\includegraphics[width=0.6\linewidth]{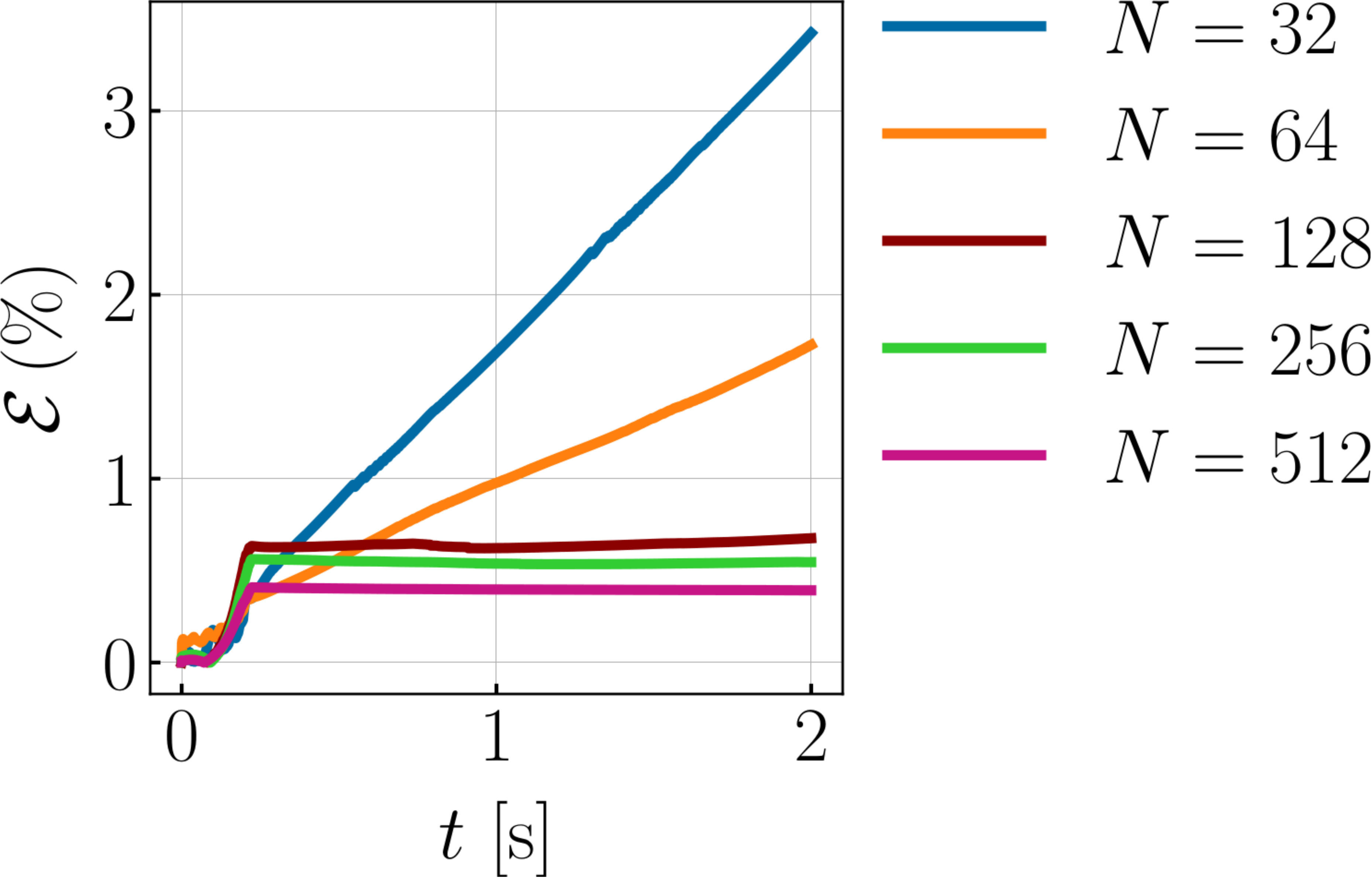}
\caption{Percentage change in PCM mass $\mathcal{E}$ as a function of time for the metal solidification case that exhibits pipe shrinkage defect at different grids.  The temperature interval is taken to be $\Delta T=10$ K for all grids. A uniform time step size is used for all the grids: for coarse grids $N=32$ and $N=64$,  $\Delta t=10^{-4}$ s is used; for medium grids $N=128$ and $N=256$, $\Delta t=10^{-5}$ s is used; and for the fine grid $N=512$, $\Delta t=10^{-6}$ s is used.}
\label{fig_pipe_shrinkage_error}
\end{figure}

For the metal melting and solidification cases discussed in the main text, we present the percentage change in the PCM mass. The mass of PCM (solid+liquid) in the domain at time $t$ can be computed using
\begin{equation}
m(t) = \int_\Omega \left[ \rhol (H\varphi) + \rhos (H-H\varphi) \right] \, \d\Omega.
\end{equation}
The relative percentage error in PCM mass is defined as
\begin{equation}
\mathcal{E}(t) = \frac{|m - m_0|}{m_0}\times 100,
\end{equation}
in which $m_0$ is the initial (at time $t=0$) mass of the PCM.  Fig.~\ref{fig_mass_conservation_error} plots $\mathcal{E}$ versus $t$ for various grids of size $N_x \times N_y = N^2$. With increasing grid refinement, $\mathcal{E}$ decreases and at finer grids there is about 0.007\% change in the PCM mass. Fig.~\ref{fig_pipe_shrinkage_error} plots $\mathcal{E}$ versus $t$ for the metal solidification case that exhibits pipe shrinkage. A similar conclusion can be drawn for this case as well: $\mathcal{E}$ decreases with grid refinement and at finer grids PCM mass changes by approximately 0.39\%. $\mathcal{E}$ also decreases with decreasing $\Delta t$ (data not shown).

%% file: stefan_manuscript.bbl
\begin{thebibliography}{10}
\expandafter\ifx\csname url\endcsname\relax
  \def\url#1{\texttt{#1}}\fi
\expandafter\ifx\csname urlprefix\endcsname\relax\def\urlprefix{URL }\fi
\expandafter\ifx\csname href\endcsname\relax
  \def\href#1#2{#2} \def\path#1{#1}\fi

\bibitem{el2017thermal}
A.~El~Khadraoui, S.~Bouadila, S.~Kooli, A.~Farhat, A.~Guizani, {Thermal
  behavior of indirect solar dryer: Nocturnal usage of solar air collector with
  PCM}, Journal of cleaner production 148 (2017) 37--48.

\bibitem{allouhi2018optimization}
A.~Allouhi, A.~A. Msaad, M.~B. Amine, R.~Saidur, M.~Mahdaoui, T.~Kousksou,
  A.~Pandey, A.~Jamil, N.~Moujibi, A.~Benbassou, {Optimization of melting and
  solidification processes of PCM: Application to integrated collector storage
  solar water heaters (ICSSWH)}, Solar Energy 171 (2018) 562--570.

\bibitem{badiei2020performance}
Z.~Badiei, M.~Eslami, K.~Jafarpur, {Performance improvements in solar flat
  plate collectors by integrating with phase change materials and fins: A CFD
  modeling}, Energy 192 (2020) 116719.

\bibitem{hossain2019two}
M.~Hossain, A.~Pandey, J.~Selvaraj, N.~Abd~Rahim, M.~Islam, V.~Tyagi, {Two side
  serpentine flow based photovoltaic-thermal-phase change materials (PVT-PCM)
  system: Energy, exergy and economic analysis}, Renewable Energy 136 (2019)
  1320--1336.

\bibitem{nie2020numerical}
C.~Nie, S.~Deng, J.~Liu, {Numerical investigation of PCM in a thermal energy
  storage unit with fins: consecutive charging and discharging}, J. Energy
  Storage 29 (2020) 101319.

\bibitem{buffo2021dynamics}
J.~Buffo, C.~Meyer, J.~Parkinson, {Dynamics of a Solidifying Icy Satellite
  Shell}, Journal of Geophysical Research: Planets 126~(5) (2021)
  e2020JE006741.

\bibitem{buffo2021characterizing}
J.~Buffo, B.~Schmidt, C.~Huber, C.~Meyer, {Characterizing the ice-ocean
  interface of icy worlds: A theoretical approach}, Icarus 360 (2021) 114318.

\bibitem{king2015laser}
W.~E. King, A.~T. Anderson, R.~M. Ferencz, N.~E. Hodge, C.~Kamath, S.~A.
  Khairallah, A.~M. Rubenchik, Laser powder bed fusion additive manufacturing
  of metals; physics, computational, and materials challenges, Applied Physics
  Reviews 2~(4) (2015) 041304.

\bibitem{king2015overview}
W.~King, A.~T. Anderson, R.~M. Ferencz, N.~E. Hodge, C.~Kamath, S.~A.
  Khairallah, {Overview of modelling and simulation of metal powder bed fusion
  process at Lawrence Livermore National Laboratory}, Materials Science and
  Technology 31~(8) (2015) 957--968.

\bibitem{khairallah2016laser}
S.~A. Khairallah, A.~T. Anderson, A.~Rubenchik, W.~E. King, {Laser powder-bed
  fusion additive manufacturing: Physics of complex melt flow and formation
  mechanisms of pores, spatter, and denudation zones}, Acta Materialia 108
  (2016) 36--45.

\bibitem{ly2017metal}
S.~Ly, A.~M. Rubenchik, S.~A. Khairallah, G.~Guss, M.~J. Matthews, Metal vapor
  micro-jet controls material redistribution in laser powder bed fusion
  additive manufacturing, Scientific reports 7~(1) (2017) 1--12.

\bibitem{kruth1996basic}
J.-P. Kruth, B.~Van~der Schueren, J.~Bonse, B.~Morren, Basic powder
  metallurgical aspects in selective metal powder sintering, CIRP annals 45~(1)
  (1996) 183--186.

\bibitem{karayagiz2019numerical}
K.~Karayagiz, A.~Elwany, G.~Tapia, B.~Franco, L.~Johnson, J.~Ma, I.~Karaman,
  R.~Arr{\'o}yave, Numerical and experimental analysis of heat distribution in
  the laser powder bed fusion of ti-6al-4v, IISE Transactions 51~(2) (2019)
  136--152.

\bibitem{heeling2017melt}
T.~Heeling, M.~Cloots, K.~Wegener, Melt pool simulation for the evaluation of
  process parameters in selective laser melting, Additive Manufacturing 14
  (2017) 116--125.

\bibitem{matthews2016denudation}
M.~J. Matthews, G.~Guss, S.~A. Khairallah, A.~M. Rubenchik, P.~J. Depond, W.~E.
  King, Denudation of metal powder layers in laser powder bed fusion processes,
  Acta Materialia 114 (2016) 33--42.

\bibitem{nangia2019robust}
N.~Nangia, B.~E. Griffith, N.~A. Patankar, A.~P.~S. Bhalla, {A robust
  incompressible Navier-Stokes solver for high density ratio multiphase flows},
  Journal of Computational Physics 390 (2019) 548--594.

\bibitem{pathak20163d}
A.~Pathak, M.~Raessi, {A 3D, fully Eulerian, VOF-based solver to study the
  interaction between two fluids and moving rigid bodies using the fictitious
  domain method}, Journal of computational physics 311 (2016) 87--113.

\bibitem{patel2018diffuse}
J.~K. Patel, G.~Natarajan, Diffuse interface immersed boundary method for
  multi-fluid flows with arbitrarily moving rigid bodies, Journal of
  Computational Physics 360 (2018) 202--228.

\bibitem{panwisawas2017mesoscale}
C.~Panwisawas, C.~Qiu, M.~J. Anderson, Y.~Sovani, R.~P. Turner, M.~M. Attallah,
  J.~W. Brooks, H.~C. Basoalto, {Mesoscale modelling of selective laser
  melting: Thermal fluid dynamics and microstructural evolution}, Computational
  Materials Science 126 (2017) 479--490.

\bibitem{wu2018numerical}
Y.-C. Wu, C.-H. San, C.-H. Chang, H.-J. Lin, R.~Marwan, S.~Baba, W.-S. Hwang,
  Numerical modeling of melt-pool behavior in selective laser melting with
  random powder distribution and experimental validation, Journal of Materials
  Processing Technology 254 (2018) 72--78.

\bibitem{wu2018parametric}
Y.-C. Wu, W.-S. Hwang, C.-H. San, C.-H. Chang, H.-J. Lin, {Parametric study of
  surface morphology for selective laser melting on Ti6Al4V powder bed with
  numerical and experimental methods}, International Journal of Material
  Forming 11~(6) (2018) 807--813.

\bibitem{gurtler2013simulation}
F.-J. G{\"u}rtler, M.~Karg, K.-H. Leitz, M.~Schmidt, Simulation of laser beam
  melting of steel powders using the three-dimensional volume of fluid method,
  Physics Procedia 41 (2013) 881--886.

\bibitem{attar2011lattice}
E.~Attar, C.~K{\"o}rner, {Lattice Boltzmann model for thermal free surface
  flows with liquid--solid phase transition}, International Journal of Heat and
  Fluid Flow 32~(1) (2011) 156--163.

\bibitem{panwisawas2017keyhole}
C.~Panwisawas, B.~Perumal, R.~M. Ward, N.~Turner, R.~P. Turner, J.~W. Brooks,
  H.~C. Basoalto, {Keyhole formation and thermal fluid flow-induced porosity
  during laser fusion welding in titanium alloys: Experimental and modelling},
  Acta Materialia 126 (2017) 251--263.

\bibitem{aggarwal2018particle}
A.~Aggarwal, A.~Kumar, {Particle scale modelling of selective laser
  melting-based additive manufacturing process using open-source CFD code
  OpenFOAM}, Transactions of the Indian Institute of Metals 71~(11) (2018)
  2813--2817.

\bibitem{cook2020simulation}
P.~S. Cook, A.~B. Murphy, {Simulation of melt pool behaviour during additive
  manufacturing: Underlying physics and progress}, Additive Manufacturing 31
  (2020) 100909.

\bibitem{wolff2019situ}
S.~J. Wolff, H.~Wu, N.~Parab, C.~Zhao, K.~F. Ehmann, T.~Sun, J.~Cao, In-situ
  high-speed x-ray imaging of piezo-driven directed energy deposition additive
  manufacturing, Scientific reports 9~(1) (2019) 1--14.

\bibitem{voller1987fixed}
V.~Voller, C.~Prakash, A fixed grid numerical modeling methodology for
  convection diffusion mushy region phase-change problem, International Journal
  Heat Mass Transfer 30~(9) (1987) 1709--1719.

\bibitem{voller1991eral}
V.~R. Voller, C.~Swaminathan, {ERAL Source-based method for solidification
  phase change}, Numerical Heat Transfer, Part B Fundamentals 19~(2) (1991)
  175--189.

\bibitem{galione2015fixed}
P.~Galione, O.~Lehmkuhl, J.~Rigola, A.~Oliva, Fixed-grid numerical modeling of
  melting and solidification using variable thermo-physical
  properties--application to the melting of n-octadecane inside a spherical
  capsule, International Journal of Heat and Mass Transfer 86 (2015) 721--743.

\bibitem{hassab2017effect}
M.~Hassab, M.~M. Sorour, M.~K. Mansour, M.~M. Zaytoun, Effect of volume
  expansion on the melting process’s thermal behavior, Applied Thermal
  Engineering 115 (2017) 350--362.

\bibitem{dallaire2017numerical}
J.~Dallaire, L.~Gosselin, Numerical modeling of solid-liquid phase change in a
  closed 2d cavity with density change, elastic wall and natural convection,
  International Journal of Heat and Mass Transfer 114 (2017) 903--914.

\bibitem{faden2019optimum}
M.~Faden, A.~K{\"o}nig-Haagen, D.~Br{\"u}ggemann, An optimum enthalpy approach
  for melting and solidification with volume change, Energies 12~(5) (2019)
  868.

\bibitem{yan2018fully}
J.~Yan, W.~Yan, S.~Lin, G.~Wagner, A fully coupled finite element formulation
  for liquid--solid--gas thermo-fluid flow with melting and solidification,
  Computer Methods in Applied Mechanics and Engineering 336 (2018) 444--470.

\bibitem{lin2020conservative}
S.~Lin, Z.~Gan, J.~Yan, G.~J. Wagner, A conservative level set method on
  unstructured meshes for modeling multiphase thermo-fluid flow in additive
  manufacturing processes, Computer Methods in Applied Mechanics and
  Engineering 372 (2020) 113348.

\bibitem{myers2020stefan}
T.~G. Myers, M.~G. Hennessy, M.~Calvo-Schwarzw{\"a}lder, {The Stefan problem
  with variable thermophysical properties and phase change temperature},
  International Journal of Heat and Mass Transfer 149 (2020) 118975.

\bibitem{delhaye1974jump}
J.-M. Delhaye, {Jump conditions and entropy sources in two-phase systems. Local
  instant formulation}, International Journal of Multiphase Flow 1~(3) (1974)
  395--409.

\bibitem{alexiades2018mathematical}
V.~Alexiades, A.~D. Solomon, Mathematical modeling of melting and freezing
  processes, Routledge, 2018.

\bibitem{huang2022consistent}
Z.~Huang, G.~Lin, A.~M. Ardekani, A consistent and conservative phase-field
  model for thermo-gas-liquid-solid flows including liquid-solid phase change,
  Journal of Computational Physics 449 (2022) 110795.

\bibitem{javierre2006comparison}
E.~Javierre, C.~Vuik, F.~Vermolen, S.~Van~der Zwaag, {A comparison of numerical
  models for one-dimensional Stefan problems}, Journal of Computational and
  Applied Mathematics 192~(2) (2006) 445--459.

\bibitem{gibou2007level}
F.~Gibou, L.~Chen, D.~Nguyen, S.~Banerjee, {A level set based sharp interface
  method for the multiphase incompressible Navier--Stokes equations with phase
  change}, Journal of Computational Physics 222~(2) (2007) 536--555.

\bibitem{khalloufi2020adaptive}
M.~Khalloufi, R.~Valette, E.~Hachem, Adaptive eulerian framework for boiling
  and evaporation, Journal of Computational Physics 401 (2020) 109030.

\bibitem{hahn2012heat}
D.~W. Hahn, M.~N. {\"O}zisik, Heat conduction, John Wiley \& Sons, 2012.

\bibitem{boettinger2002phase}
W.~J. Boettinger, J.~A. Warren, C.~Beckermann, A.~Karma, Phase-field simulation
  of solidification, Annual review of materials research 32~(1) (2002)
  163--194.

\bibitem{hu1996mathematical}
H.~Hu, S.~A. Argyropoulos, Mathematical modelling of solidification and
  melting: a review, Modelling and Simulation in Materials Science and
  Engineering 4~(4) (1996) 371.

\bibitem{moukalled2016finite}
F.~Moukalled, L.~Mangani, M.~Darwish, The finite volume method in computational
  fluid dynamics, Vol. 113, Springer, 2016.

\bibitem{patankar2018numerical}
S.~Patankar, Numerical heat transfer and fluid flow, Taylor \& Francis, 2018.

\bibitem{Tan09}
F.~Tan, S.~Hosseinizadeh, J.~Khodadadi, L.~Fan, {Experimental and computational
  study of constrained melting of phase change materials (PCM) inside a
  spherical capsule}, International Journal of Heat and Mass Transfer
  52~(15-16) (2009) 3464--3472.

\bibitem{Beckermann88}
C.~Beckermann, R.~Viskanta, Natural convection solid/liquid phase change in
  porous media, International journal of heat and mass transfer 31~(1) (1988)
  35--46.

\bibitem{pember1998adaptive}
W.~Fiveland, J.~Jessee, An adaptive projection method for unsteady, low-mach
  number combustion, Combustion Science and Technology 140~(1-6) (1998)
  123--168.

\bibitem{hosseini2022low}
S.~A. Hosseini, N.~Darabiha, D.~Th{\'e}venin, Low mach number lattice boltzmann
  model for turbulent combustion: flow in confined geometries, Proceedings of
  the Combustion Institute (2022).

\bibitem{bell2004adaptive}
J.~Bell, M.~Day, C.~Rendleman, S.~Woosley, M.~Zingale, {Adaptive low Mach
  number simulations of nuclear flame microphysics}, Journal of Computational
  Physics 195~(2) (2004) 677--694.

\bibitem{gilet2013low}
C.~Gilet, A.~Almgren, J.~Bell, A.~Nonaka, S.~Woosley, M.~Zingale, Low mach
  number modeling of core convection in massive stars, The Astrophysical
  Journal 773~(2) (2013) 137.

\bibitem{donev2014low}
A.~Donev, A.~Nonaka, Y.~Sun, T.~Fai, A.~Garcia, J.~Bell, {Low mach number
  fluctuating hydrodynamics of diffusively mixing fluids}, Communications in
  Applied Mathematics and Computational Science 9~(1) (2014) 47--105.

\bibitem{donev2015low}
A.~Donev, A.~Nonaka, A.~K. Bhattacharjee, A.~L. Garcia, J.~B. Bell, Low mach
  number fluctuating hydrodynamics of multispecies liquid mixtures, Physics of
  Fluids 27~(3) (2015) 037103.

\bibitem{nonaka2015low}
A.~Nonaka, Y.~Sun, J.~Bell, A.~Donev, Low mach number fluctuating hydrodynamics
  of binary liquid mixtures, Communications in Applied Mathematics and
  Computational Science 10~(2) (2015) 163--204.

\bibitem{stefanescu2015science}
D.~M. Stefanescu, Science and engineering of casting solidification, Springer,
  2015.

\bibitem{alexiades1993weak}
V.~Alexiades, J.~Drake, A weak formulation for phase-change problems with bulk
  movement due to unequal densities, PITMAN RESEARCH NOTES IN MATHEMATICS
  SERIES (1993) 82--82.

\bibitem{durlofsky1987analysis}
L.~Durlofsky, J.~Brady, {Analysis of the Brinkman equation as a model for flow
  in porous media}, The Physics of fluids 30~(11) (1987) 3329--3341.

\bibitem{Sussman1994}
M.~Sussman, P.~Smereka, S.~Osher, A level set approach for computing solutions
  to incompressible two-phase flow, Journal of Computational Physics 114~(1)
  (1994) 146--159.

\bibitem{brackbill1992continuum}
J.~U. Brackbill, D.~B. Kothe, C.~Zemach, A continuum method for modeling
  surface tension, Journal of computational physics 100~(2) (1992) 335--354.

\bibitem{saldi2012marangoni}
Z.~Saldi, Marangoni driven free surface flows in liquid weld pools, Ph.D.
  thesis (2012).

\bibitem{francois2006balanced}
M.~M. Francois, S.~J. Cummins, E.~D. Dendy, D.~B. Kothe, J.~M. Sicilian, M.~W.
  Williams, A balanced-force algorithm for continuous and sharp interfacial
  surface tension models within a volume tracking framework, Journal of
  Computational Physics 213~(1) (2006) 141--173.

\bibitem{thirumalaisamy2023pre}
R.~Thirumalaisamy, K.~Khedkar, P.~Ghysels, A.~P.~S. Bhalla,
  \href{https://www.sciencedirect.com/science/article/pii/S0021999123004205}{An
  effective preconditioning strategy for volume penalized incompressible/low
  mach multiphase flow solvers}, Journal of Computational Physics 490 (2023)
  112325.
\newblock \href {https://doi.org/https://doi.org/10.1016/j.jcp.2023.112325}
  {\path{doi:https://doi.org/10.1016/j.jcp.2023.112325}}.
\newline\urlprefix\url{https://www.sciencedirect.com/science/article/pii/S0021999123004205}

\bibitem{Nangia2019WSI}
N.~Nangia, N.~A. Patankar, A.~P.~S. Bhalla, {A DLM immersed boundary method
  based wave-structure interaction solver for high density ratio multiphase
  flows}, Journal of Computational Physics 398 (2019) 108804.

\bibitem{IBAMR-web-page}
{IBAMR}: {A}n adaptive and distributed-memory parallel implementation of the
  immersed boundary method, \url{https://github.com/IBAMR/IBAMR}.

\bibitem{HornungKohn02}
R.~D. Hornung, S.~R. Kohn, Managing application complexity in the {SAMRAI}
  object-oriented framework, Concurrency Comput Pract Ex 14~(5) (2002)
  347--368.

\bibitem{samrai-web-page}
{SAMRAI}: {S}tructured {A}daptive {M}esh {R}efinement {A}pplication
  {I}nfrastructure, \url{http://www.llnl.gov/CASC/SAMRAI}.

\bibitem{petsc-user-ref}
S.~Balay, S.~Abhyankar, M.~F. Adams, J.~Brown, P.~Brune, K.~Buschelman,
  L.~Dalcin, V.~Eijkhout, W.~D. Gropp, D.~Kaushik, M.~G. Knepley, L.~C.
  McInnes, K.~Rupp, B.~F. Smith, S.~Zampini, H.~Zhang,
  \href{http://www.mcs.anl.gov/petsc}{{PETS}c users manual}, Tech. Rep.
  ANL-95/11 - Revision 3.6, Argonne National Laboratory (2015).
\newline\urlprefix\url{http://www.mcs.anl.gov/petsc}

\bibitem{petsc-web-page}
S.~Balay, S.~Abhyankar, M.~F. Adams, J.~Brown, P.~Brune, K.~Buschelman,
  L.~Dalcin, V.~Eijkhout, W.~D. Gropp, D.~Kaushik, M.~G. Knepley, L.~C.
  McInnes, K.~Rupp, B.~F. Smith, S.~Zampini, H.~Zhang,
  \href{http://www.mcs.anl.gov/petsc}{{PETS}c {W}eb page},
  \url{http://www.mcs.anl.gov/petsc} (2015).
\newline\urlprefix\url{http://www.mcs.anl.gov/petsc}

\bibitem{childs2012visit}
H.~Childs, E.~Brugger, B.~Whitlock, J.~Meredith, S.~Ahern, D.~Pugmire,
  K.~Biagas, M.~Miller, C.~Harrison, G.~H. Weber, et~al., {VisIt: An end-user
  tool for visualizing and analyzing very large data}, Tech. rep. (2012).

\end{thebibliography}
